\ifdefined\isabridged 
\documentclass[letterpaper,twocolumn,10pt]{article}
\usepackage{usenix,epsfig,endnotes}
\else
\documentclass[sigconf]{acmart}
\fi
\usepackage[T1]{fontenc}
\usepackage[utf8]{inputenc}
\usepackage{letltxmacro}
\usepackage{todonotes}
\usepackage{xspace}
\usepackage{hyperref}
\usepackage{lmodern}
\usepackage{pifont}
\usepackage{listings}
\usepackage{enumitem}
\usepackage{multirow}
\usepackage{amsmath}
\usepackage{amssymb}
\usepackage{times}
\usepackage{graphics}
\usepackage{microtype}

\ifdefined\isabridged 
\usepackage[numbers]{natbib}
\fi

\newif\ifabridged
\newif\ifnotabridged
\newif\ifanonymous
\newif\ifnotanonymous

\ifdefined\isabridged
\abridgedtrue
\fi

\ifabridged\notabridgedfalse
\else\notabridgedtrue
\fi

\ifdefined\isanonymous
\anonymoustrue
\fi

\ifanonymous\notanonymousfalse
\else\notanonymoustrue
\fi

\LetLtxMacro{\todonote}{\todo}
\renewcommand{\todo}[2][]
{\todonote[inline, caption={#2}, size=\footnotesize, #1]
{\renewcommand{\baselinestretch}{0.5}\selectfont#2\par}}

\ifanonymous
\DeclareRobustCommand\sectt[1]{{\fontsize{13}{12}#1}}

\newcommand{\LONGNAME}{\protect{Run-time Scope Enforcement}\xspace}

\newcommand{\ABBRNAME}{\protect{RSE}\xspace}
\newcommand{\SHORTNAMETITLE}{\protect\sectt{XSCEN}\xspace}
\newcommand{\SHORTNAME}{\protect{\scshape{XSCEN}}\xspace}
\newcommand{\SHORTNAMEDESC}{\protect{Extensible Scope Enforcement}\xspace}
\else
\DeclareRobustCommand\sectt[1]{{\fontsize{13}{12}#1}}

\newcommand{\LONGNAME}{\protect{Run-time Scope Enforcement}\xspace}

\newcommand{\ABBRNAME}{\protect{RSE}\xspace}
\newcommand{\SHORTNAMETITLE}{\protect\sectt{HardScope}\xspace}
\newcommand{\SHORTNAME}{\protect{\textsf{HardScope}}\xspace}
\fi

\newcommand{\dOne}{\ding{182}}
\newcommand{\dTwo}{\ding{183}}
\newcommand{\dThree}{\ding{184}}
\newcommand{\dFour}{\ding{185}}
\newcommand{\dFive}{\ding{186}}
\newcommand{\dSix}{\ding{187}}
\newcommand{\dSeven}{\ding{188}}
\newcommand{\dEight}{\ding{189}}
\newcommand{\dNine}{\ding{190}}
\newcommand{\dTen}{\ding{191}}

\ifanonymous
\newcommand{\materials}{\url{https://goo.gl/TAjLxy}}

\else
\newcommand{\materials}{\url{https://goo.gl/TAjLxy}}
\fi

\newcommand{\myparagraph}[1]{\textbf{#1.}}

\newcommand{\mylistingwidth}{.95\columnwidth}

\newcommand{\mytablefootnotewidth}{.95\columnwidth}

\newcommand{\myappxsection}[1]{\section{#1}}

\lstdefinestyle{customc}{
  belowcaptionskip=1\baselineskip,
  breaklines=true,
  frame=L,
  xleftmargin=\parindent,
  language=C,
  showstringspaces=false,
  basicstyle=\footnotesize\ttfamily,
  keywordstyle=\bfseries\color{green!40!black},
  commentstyle=\itshape\color{purple!40!black},
  identifierstyle=\color{blue},
  stringstyle=\color{orange},
}

\lstdefinelanguage[riscvasm]{Assembler}{morekeywords={,lui,auipc,jal,jalr,beq,bne,blt,bge,bltu,bgeu,lb,lh,lw,    lbu,lhu,sb,sh,sw,addi,slti,sltiu,xori,ori,andi,slli,srli,srai,add,    sub,sll,slt,sltu,xor,srl,sra,or,and,fence,fence.i,ecall,ebreak,csrrw,    csrrs,csrrc,csrrwi,csrrsi,csrrci,lwu,ld,sd,addiw,slliw,srliw,sraiw,    addw,subw,sllw,srlw,sraw,,mul,mulh,mulhu,mulhsu,div,divu,rem,remu,    mulw,divw,divuw,remw,remuw,,lr.w,sc.w,amoswap.w,amoadd.w,amoxor.w,    amoand.w,amoor.w,amomin.w,amomax.w,amominu.w,amomaxu.w,,lr.d,sc.d,    amoswap.d,amoadd.d,amoxor.d,amoand.d,amoor.d,amomin.d,amomax.d,    amominu.d,amomaxu.d,,flw,fsw,fmadd.s,fmsub.s,fnmsub.s,fnmadd.s,fadd.s,    fsub.s,fmul.s,fdiv.s,fsqrt.s,fsgnj.s,fsgnjn.s,fsgnjx.s,fmin.s,fmax.s,    fcvt.w.s,fcvt.wu.s,fmv.x.s,feq.s,flt.s,fle.s,fclass.s,fcvt.s.w,fcvt.s.wu,    fmv.s.x,,fcvt.l.s,fcvt.lu.s,fcvy.s.l,fcvt.s.lu,,fld,fsd,fmadd.d,fmsub.d,    fnmsub.d,fnmadd.d,fadd.d,fsub.d,fmul.d,fdiv.d,fsqrt.d,fsgnj.d,fsgnjn.d,    fsgnjx.d,fmin.d,fmax.d,fcvt.s.d,fcvt.d.s,feq.d,flt.d,fle.d,fclass.d,    fcvt.w.d,fcvt.wu.d,fcvt.d.w,fcvt.d.wu,,fcvt.l.d,fcvt.lu.d,fmv.x.d,fcvt.d.l,    fcvt.d.lu,fmv.d.x,,c.addi4spn,c.fld,c.lq,c.lw,c.flw,c.ld,c.fsd,c.sq,c.sw,    c.fsw,c.sd,c.nop,c.addi,c.jal,c.addiw,c.li,c.addi16sp,c.lui,c.srli,c.srli64,    c.srai,c.srai64,c.andi,c.sub,c.xor,c.or,c.and,c.subw,c.addw,c.j,c.beqz,    c.bnez,c.slli,c.slli64,c.fldsp,c.lqsp,c.lwsp,c.flwsp,c.ldsp,c.jr,c.mv,    c.ebreak,c.jalr,c.add,c.fsdsp,c.sqsp,c.fswsp,c.sdsp,,getq,setq,retirq,    maskirq,waitirq,timer,,la,nop,li,mv,not,neg,negw,sext.w,seqz,snez,sltz,    sgtz,fmv.s,fabs.s,fneg.s,fmv.d,fabs.d,fneg.d,beqz,bnez,blez,bgez,bltz,    bgtz,j,jr,ret,call,tail,rdtimer,rdtimerh,rdcycle,rdcycleh,rdinstr,rdinstrh}   
    alsoletter=.,alsodigit=?,  sensitive=f,  morestring=[b]",  morestring=[b]',  morecomment=[l];}[keywords,comments,strings]

\lstdefinestyle{customasm}{
  belowcaptionskip=1\baselineskip,
  frame=L,
  xleftmargin=\parindent,
  language=[riscvasm]Assembler,
  basicstyle=\footnotesize\ttfamily,
  commentstyle=\itshape\color{purple!40!black},
}

\ifnotabridged                    \renewcommand\footnotetextcopyrightpermission[1]{}                                   
\begin{abstract}

  Widespread use of memory unsafe programming languages (e.g., C and C++)
    leaves many systems vulnerable to \emph{memory corruption attacks}.
  A variety of defenses have been proposed to mitigate attacks that exploit memory errors to hijack the control flow of the code at run-time, e.g., (fine-grained) randomization or Control Flow Integrity.   
However, recent work on \emph{data-oriented programming} (DOP) demonstrated highly expressive (Turing-complete) attacks, even in the presence of these state-of-the-art defenses.
Although multiple real-world DOP attacks have been demonstrated, no efficient defenses are yet available.
  We propose \emph{run-time scope enforcement} (\ABBRNAME), a novel approach designed to \emph{efficiently mitigate all currently known DOP attacks} by enforcing compile-time memory safety constraints (e.g., variable visibility rules) at run-time. We present \SHORTNAME, a proof-of-concept implementation of hardware-assisted \ABBRNAME for the new RISC-V open instruction set architecture.
  We \ifnotabridged discuss our systematic empirical evaluation of \SHORTNAME which demonstrates \else demonstrate \fi that it can mitigate all currently known DOP attacks, and has a real-world performance overhead of 3.2\% in embedded benchmarks.
\end{abstract}

  \fi

\ifnotanonymous 
\ifabridged  \author{
\todo{TODO: USENIX author block}
}
\else  \author{Thomas Nyman}
\affiliation{  \institution{Aalto University, Finland}
}
\email{thomas.nyman@aalto.fi}

\author{Ghada Dessouky}
\affiliation{  \institution{Technische Universit\"at Darmstadt, Germany}
}
\email{ghada.dessouky@trust.tu-darmstadt.de}

\author{Shaza Zeitouni}
\affiliation{  \institution{Technische Universit\"at Darmstadt, Germany}
}
\email{shaza.zeitouni@trust.tu-darmstadt.de}

\author{Aaro Lehikoinen}
\affiliation{  \institution{Aalto University, Finland}
}
\email{aaro.j.lehikoinen@aalto.fi}

\author{Andrew Paverd}
\affiliation{  \institution{Aalto University, Finland}
}
\email{andrew.paverd@ieee.org}

\author{N. Asokan}
\affiliation{  \institution{Aalto University, Finland}
}
\email{asokan@acm.org}

\author{Ahmad-Reza Sadeghi}
\affiliation{  \institution{Technische Universit\"at Darmstadt, Germany}
}
\email{ahmad.sadeghi@trust.tu-darmstadt.de}

\renewcommand{\shortauthors}{T. Nyman et al.}
\fi
\fi

\begin{document}

\ifanonymous
\title{\SHORTNAMETITLE: \SHORTNAMEDESC}
\else
\title{\SHORTNAMETITLE: Thwarting DOP attacks with Hardware-assisted Run-time Scope Enforcement}
\fi

\maketitle

\ifnotabridged
\hfill
\fi

\ifabridged
\begin{abstract}

  Widespread use of memory unsafe programming languages (e.g., C and C++)
    leaves many systems vulnerable to \emph{memory corruption attacks}.
  A variety of defenses have been proposed to mitigate attacks that exploit memory errors to hijack the control flow of the code at run-time, e.g., (fine-grained) randomization or Control Flow Integrity.   
However, recent work on \emph{data-oriented programming} (DOP) demonstrated highly expressive (Turing-complete) attacks, even in the presence of these state-of-the-art defenses.
Although multiple real-world DOP attacks have been demonstrated, no efficient defenses are yet available.
  We propose \emph{run-time scope enforcement} (\ABBRNAME), a novel approach designed to \emph{efficiently mitigate all currently known DOP attacks} by enforcing compile-time memory safety constraints (e.g., variable visibility rules) at run-time. We present \SHORTNAME, a proof-of-concept implementation of hardware-assisted \ABBRNAME for the new RISC-V open instruction set architecture.
  We \ifnotabridged discuss our systematic empirical evaluation of \SHORTNAME which demonstrates \else demonstrate \fi that it can mitigate all currently known DOP attacks, and has a real-world performance overhead of 3.2\% in embedded benchmarks.
\end{abstract}

\fi

\section{Introduction}
\label{sec:introduction}

Although known for over two decades, memory corruption vulnerabilities are still a persistent source of threats against software systems.
The main problem is that modern software still contains a lot of unsafe code written in memory unsafe programming languages (e.g., C and C++), especially in embedded systems and the Internet of Things~\cite{IoTDeveloperSurvey17}.
The lack of memory safety in these languages and the inevitability of software bugs leave many systems vulnerable to attacks that exploit \emph{memory errors}.

Control-flow attacks, such as \emph{Return-Oriented Programming} (ROP)~\cite{Shacham07}, which hijack the execution flow of a program, are well-known, and various defenses against them have been proposed (e.g.,~\citep{Abadi09,Kuznetsov14,Larsen14}).
In contrast, \emph{non-control-data} attacks do not need to modify the control-flow, and thus cannot be prevented by these defenses.
Instead, non-control-data attacks corrupt data used for decision-making, e.g., to leak sensitive data or escalate privileges by corrupting variables used in authorization decisions.
Some defenses against non-control-data attacks (e.g.,~\citep{Castro09,Schlesinger11}) provide protection against attacks that only target individual pieces of (security-critical) data.

However, recent work has shown that non-control-data attacks can be generalized to achieve Turing-complete execution, called \emph{Data-Orientated Programming} (DOP)~\cite{Hu16}.
In DOP, the attacker carefully corrupts non-control-data to build up sequences of operations (\emph{data-oriented gadgets}) without modifying the program's control-flow.
Each gadget simulates a virtual operation on some attacker-controlled input.
Unlike previous non-control-data attacks, DOP can be highly expressive (e.g., including assignment, arithmetic, and conditional decisions).
This allows DOP to actively break state-of-the-art defenses, such as \emph{Address Space Layout Randomization} (ASLR)~\cite{Larsen14}.
Since DOP can reuse virtually any data, preventing DOP is a significant challenge.

Practical DOP attacks have already been shown against real-world software~\cite{Hu16,Evans16}.
Hu et al.~\cite{Hu16} discuss various existing schemes that could reduce the number of DOP attacks, including memory safety, data-flow integrity, fine-grained data-plane randomization, and hardware/software fault isolation.
However, they explain that mitigating DOP with existing approaches results in high performance overheads, and do not offer viable alternatives.
As defenses against control-flow attacks become widespread~\cite{Intel-CET}, DOP is likely to become the next appealing attack technique for modern run-time exploitation.

\myparagraph{Goals and Contributions} We propose a new \emph{efficient} defense against non-control-data attacks and \emph{all currently known} DOP attacks.
The intuition behind our approach is simple: In \emph{block structured languages}, such as C and C++, every variable has a so-called \emph{lexical scope}, denoting the block(s) of source code in which the variable is visible.
Developers can thus define variables with limited \emph{scope} of visibility (e.g., local variables).
All correct compilers enforce \emph{variable scope} at compile-time by checking these variable visibility rules.
However, all currently known DOP attacks violate variable scope rules at run-time, since there is no equivalent enforcement.
Consequently, mechanisms for variable scope enforcement \emph{at run-time} can significantly reduce the number of available DOP gadgets.

In this paper, we define the notion of \emph{\LONGNAME} (\ABBRNAME) that provides \emph{fine-grained compartmentalization} of data memory within programs.
We demonstrate that \ABBRNAME can mitigate all currently known DOP attacks.
We stress that while it is not possible to guarantee the absence of DOP gadgets in arbitrary programs\footnote{A pathological case would be a program that contains all necessary gadgets, a gadget dispatcher, and all the data in the same function.} we argue that \ABBRNAME can prevent DOP in typical programs.
Unlike \ABBRNAME, existing defences (a) do not provide \emph{complete mediation} of all variable accesses~\cite{Hu16} (as we explain in Section~\ref{sec:related-work}) and (b) suffer from high performance and memory overhead~\cite{Nagarakatte09,Nagarakatte10,Castro06}.

We then describe \SHORTNAME \ifanonymous{ (\SHORTNAMEDESC)\fi, a \emph{hardware-assisted} \ABBRNAME scheme.
\ifnotabridged\SHORTNAME introduces a set of six new instructions.
Compiler-assisted instrumentation places \SHORTNAME instructions in the program to ensure that all memory access constraints are also enforced \emph{at run-time}.
As the program executes, these instructions dynamically create \emph{rules} that define which code blocks can access which pieces of memory.
One significant challenge is to minimize the performance overhead of checking these rules on every memory load/store operation.
To overcome this challenge, we designed an efficient method for storing the rules as a stack, such that all rules applicable to the currently executing code block are always at the top of the stack and can be checked simultaneously (Section~\ref{sec:implementation}).

Since enforcement rules are created and updated dynamically,\fi \SHORTNAME (or any other \ABBRNAME scheme) enables \emph{context-specific} memory isolation, unlike prior defenses that only allow static policies~\cite{Wahbe93a,Erlingsson06,Castro06,Akritidis08,Bhatkar08,Cadar08}.
This means that the same piece of code can be granted access to different memory locations depending on the context in which the code is executed.
For example, if a particular function can be called as part of either a privileged or unprivileged execution path, \SHORTNAME can allow/deny it access to certain variables in memory depending on which path was executed.\footnote{\ABBRNAME can in fact enforce different policies on each distinct invocation of a given function even in a single execution path.}
This is critical to reducing the number of available DOP gadgets.
In Section~\ref{sec:instrumentation} we describe how \SHORTNAME can enforce memory protection at either coarser or finer granularity. 
\ABBRNAME can be instantiated with \SHORTNAME to enforce various protection models, such as defending against ROP attacks by protecting function return addresses on the stack.

We developed a proof-of-concept implementation of \SHORTNAME targeting the RISC-V instruction set architecture. 
We integrated \SHORTNAME with the open-source Pulpino core\footnote{\url{http://www.pulp-platform.org/}} on a Zedboard FPGA\footnote{\url{http://zedboard.org/}} (Section~\ref{sec:implementation}). 
We also added support for \SHORTNAME to the RISC-V toolchain and enhanced the RISC-V compiler to \emph{automatically} instrument programs to protect variables at run-time to mitigate known DOP attacks. 
Finally, we extended the official RISC-V simulator, \emph{Spike}, to support our new instructions.
Our evaluation indicates that \SHORTNAME is efficient enough to be realized even on small embedded devices.

In summary, our main contributions are as follows:
\begin{itemize}
  \item \emph{\LONGNAME}: A novel approach for fine-grained \textbf{context-specific} \textbf{memory isolation} within programs (Section~\ref{sec:design}) to defeat non-control-data attacks, and all currently known DOP attacks.
  \item \SHORTNAME: An open-source proof-of-concept implementation of hardware-assisted runtime scope enforcement on the RISC-V instruction set architecture to achieve \textbf{efficient compartmentalization} of memory accesses within programs capable of mitigating all currently known DOP attacks (Section~\ref{sec:implementation}).
  \item \emph{Automatic Instrumentation}: Compiler support for \textbf{protecting large classes of variables at run-time} without requiring any developer input or data-flow analysis (Section~\ref{sec:instrumentation}).
  \item \emph{Evaluation}: Systematic analyses of how \SHORTNAME mitigates published DOP attacks (Section~\ref{subsec:dop_mitigation}), discussion of \ABBRNAME security guarantees (Section~\ref{subsec:security_considerations}), and evaluation of \SHORTNAME's hardware area overhead and performance impact (Section~\ref{subsec:performance_evaluation}). \SHORTNAME is efficient, incurring only a 3.2\% performance overhead in embedded benchmarks.
\end{itemize}

\noindent
\myparagraph{Code Availability} To enable our results to be reproduced, and to encourage further research in this area, the source code for the \SHORTNAME-enhanced GCC toolchain including the RISC-V simulator are available in the accompanying supplementary material at \materials.

\section{Background}
\label{sec:background}
\label{sec:problem-statement}

\ifnotabridged
\myparagraph{Memory errors and bounds checking} Exploitable memory errors may be used as entry-points to a vulnerable program. These may provide read or write access to program memory. At run-time, C and C++
programs can access or overwrite data anywhere in their own memory space. Modern compilers may insert checks around operations on local, global, and heap objects (e.g., arrays), that verify at run-time whether data written to a memory object is within the boundaries of that object. Such bounds checking can prevent buffer overflows, but requires instrumentation of code and incurs high performance and memory overhead~\cite{Serebryany12}, even with hardware support for such bounds checks~\cite{Oleksenko17}\footnotemark. 

\footnotetext{Major compiler collections, such as GCC and LLVM, already support run-time bounds checks using AddressSanitizer~\cite{Serebryany12} and Intel's Memory Protection Extensions (MPX)~\cite{Oleksenko17}.}

\myparagraph{Existing defenses} Modern systems prevent direct modification of program code by W$\oplus$X memory access policies such as Data Execution Prevention~\cite{HP-DEP}, use control-flow integrity~\cite{Abadi09,Intel-CET} to deter control-flow attacks that do not modify program code and raise the bar against run-time attacks using Address Space Layout Randomization (ASLR)~\cite{Larsen14}.

\myparagraph{W$\oplus$X} An attacker can use memory vulnerabilities to subvert the integrity of the program memory. Direct modification of program code in modern processors is prevented by W$\oplus$X memory access policies such as Data Execution Prevention~\cite{HP-DEP}. 
\myparagraph{Probabilistic defenses} In sophisticated run-time attacks, the attacker crafts payloads that cause the program to behave in an unintended manner. These payloads usually refer to data and code by their addresses in memory. Attackers can find these addresses by offline analysis of the program memory layout. Address Space Layout Randomization (ASLR)~\cite{Larsen14} randomizes the memory layout of the program on each execution. ASLR typically randomizes the base address of the executable and the positions of the stack, heap, and libraries. This prevents an attacker from reliably addressing known targets, for instance identifying a particular function to jump to, or reading/modifying a particular variable. However, ASLR defenses are susceptible to information leakage (e.g., by obtaining the value of a well-known pointer), and are routinely bypassed in real-world exploits~\cite{Shacham04}. 

\myparagraph{Data-Orientated Programming} \fi
In DOP~\cite{Hu16}, an attacker carefully tampers with non-control-data to execute sequences of operations within the program on attacker-controlled input.
Each sequence of operations constitutes a \emph{data-oriented gadget} which represents a single virtual machine instruction executing on top of benign program logic.
A \emph{gadget dispatcher} (e.g., an attacker-controlled loop in the benign program) enables the attacker to chain together data-orientated gadgets to realize expressive computation.

Hu et al.~\cite{Hu16} demonstrate three practical DOP attacks against the ProFTPD file server and one DOP attack against the Wireshark network packet analyzer. In the following, we describe the first attack against ProFTPD in detail. The remaining DOP attacks described by Hu et al.\ and an independently discovered attack by Evans~\cite{Evans16} against the GStreamer FLIC decoder are described in Appendix~\ref{appx:otherdop}. 

Each attack against ProFTPD exploits the same stack buffer overflow vulnerability in a general-purpose string replacement function, \texttt{sreplace()}\footnote{CVE-2006-5815: \url{https://cve.mitre.org/cgi-bin/cvename.cgi?name=cve-2006-5815}} allowing the attacker to read and write arbitrary memory locations (as shown in Appendix~\ref{appx:sreplace}). 
The attacker's goal is to obtain the server's OpenSSL private key, but the memory address of this key is randomized by the program. 
The attacker constructs a virtual DOP program that accesses the OpenSSL context structure (\texttt{ssl\_ctx}) from a well-known location in memory, then dereferences a chain of pointers to determine the key's location.
The attack requires three different data-oriented gadgets: \emph{assignment}, \emph{addition}, and \emph{pointer dereferencing}. 
The assignment gadget is constructed from the vulnerable \texttt{sreplace()} function. 
The addition gadget is realized by corrupting two integer fields in a global data structure, \texttt{session.total\_bytes\_out} and \texttt{session.xfer.total\_out}, and performs the operation \texttt{session.total\_bytes\_out += session.xfer.total\_out}. 
The dereference gadget is obtained by corrupting a string pointer in another global data structure, \texttt{main\_server.ServerName}. 
This dereferences the pointer \texttt{main\_server->ServerName} and copies the result to a known position in a static buffer. 
These gadgets can be triggered in arbitrary sequences using specially crafted input \emph{without compromising the control-flow of ProFTPD}. 
Note that during benign execution, \texttt{sreplace()} need not access \texttt{ssl\_ctx}, nor any of the nested structures that lead to the key, nor the key itself. 
In Section~\ref{subsec:dop_mitigation} we demonstrate how enforcing \ABBRNAME in ProFTPD thwarts this attack.

\section{Requirements \& Assumptions}
\label{sec:adversary-model}
\label{sec:requirements}
\label{subsec:adversary-model} \label{subsec:requirements}    \label{sec:assumptions}

\myparagraph{Adversary Model} We consider a powerful adversary who has full control over the data memory of the target program. 
This models buffer overflows and other memory corruption vulnerabilities (e.g., an externally controlled format string\footnote{CWE-134: Use of Externally-Controlled Format String \url{https://cwe.mitre.org/data/definitions/134.html}}) that could lead to arbitrary corruption of data memory. 
However, the adversary cannot modify program code at run-time (W$\oplus$X protection). 
Our adversary model is standard for runtime attacks and consistent with the adversary in Hu et al.'s DOP attacks against ProFTPD (Section~\ref{sec:background} and Appendix~\ref{appx:otherdop}).

\myparagraph{Requirements} We require a mechanism that prevents the above adversary from mounting DOP attacks.
Since DOP attacks require the adversary to modify data in unintended ways at run-time, these attacks can be prevented by a \emph{run-time enforcement mechanism} that prevents any modification of control-data and non-control-data that would not be permitted during a compile-time check by a correct compiler.
We derive the following requirements for a mechanism that mitigates all currently known DOP attacks.

\begin{description}[style=unboxed,leftmargin=0cm]

\item [R1. Multi-granularity enforcement.] Enforce memory protection at run-time for any granularity of protection domain (subject) and protected region (object).

\item [R2. Context-specific enforcement.] Enforce different permissions on each invocation of the same subject (e.g., each function), to minimize the attack surface following the principle of \emph{least privilege}.

\item [R3. Complete mediation.] Protection domains cannot increase their permissions accidentally or maliciously, and all memory accesses can be checked with only minimal performance impact and memory overhead. 

\end{description}

\noindent
\myparagraph{Design goals} We define two design goals for the system:

\begin{itemize}[noitemsep,topsep=0pt,after=\vspace{.5\baselineskip}]

\item Legacy software should run without recompilation even if selected components, such as libraries, make use of fine-grained protection.

\item Performance and memory overhead should scale gracefully with the number of protection domains (subjects), the number of protected regions (objects), and the frequency of domain transitions.

\end{itemize}

\noindent
\myparagraph{Assumptions} In our implementation of \ABBRNAME we make the following assumptions:

\begin{itemize}[noitemsep,nolistsep]

\item We restrict our attention to single-threaded C programs. 
We outline what would be needed to relax this assumption in Appendix~\ref{appx:extensions}.

\item Typical programs minimize the scope of variables and interdependence between modules, e.g., local and static variables are preferred over global variables. 
We discuss narrowing the run-time visibility of global variables in Section~\ref{sec:instrumentation}.

\item Typical programs enhance spatial locality by nesting structures instead of creating links between separate structures by nesting pointers. 
This is reasonable to assume because it improves performance by making better use of processor caches, and may also improve power consumption in embedded applications~\cite{Ibrahim13}.
Nevertheless, we also discuss how \ABBRNAME can be applied to nested pointers in Section~\ref{sec:instrumentation}.

\item We focus on an adversary that employs DOP and other non-control-data attacks. 
A real-world adversary may also attempt to influence a program's control flow. 
Defenses against control-flow attacks, such as \emph{Control-Flow Integrity} (CFI)~\cite{Abadi09} are complementary to \ABBRNAME.
In Appendix~\ref{appx:ra-protection} we show how applying \ABBRNAME at a suitable granularity can also prevent large classes of control-flow attacks, e.g., ROP.

\end{itemize}

\section{Design Overview}
\label{sec:design}

\begin{figure*}
\centering
\begin{minipage}{\columnwidth}
\begin{center}
  \includegraphics[width=0.85\columnwidth]{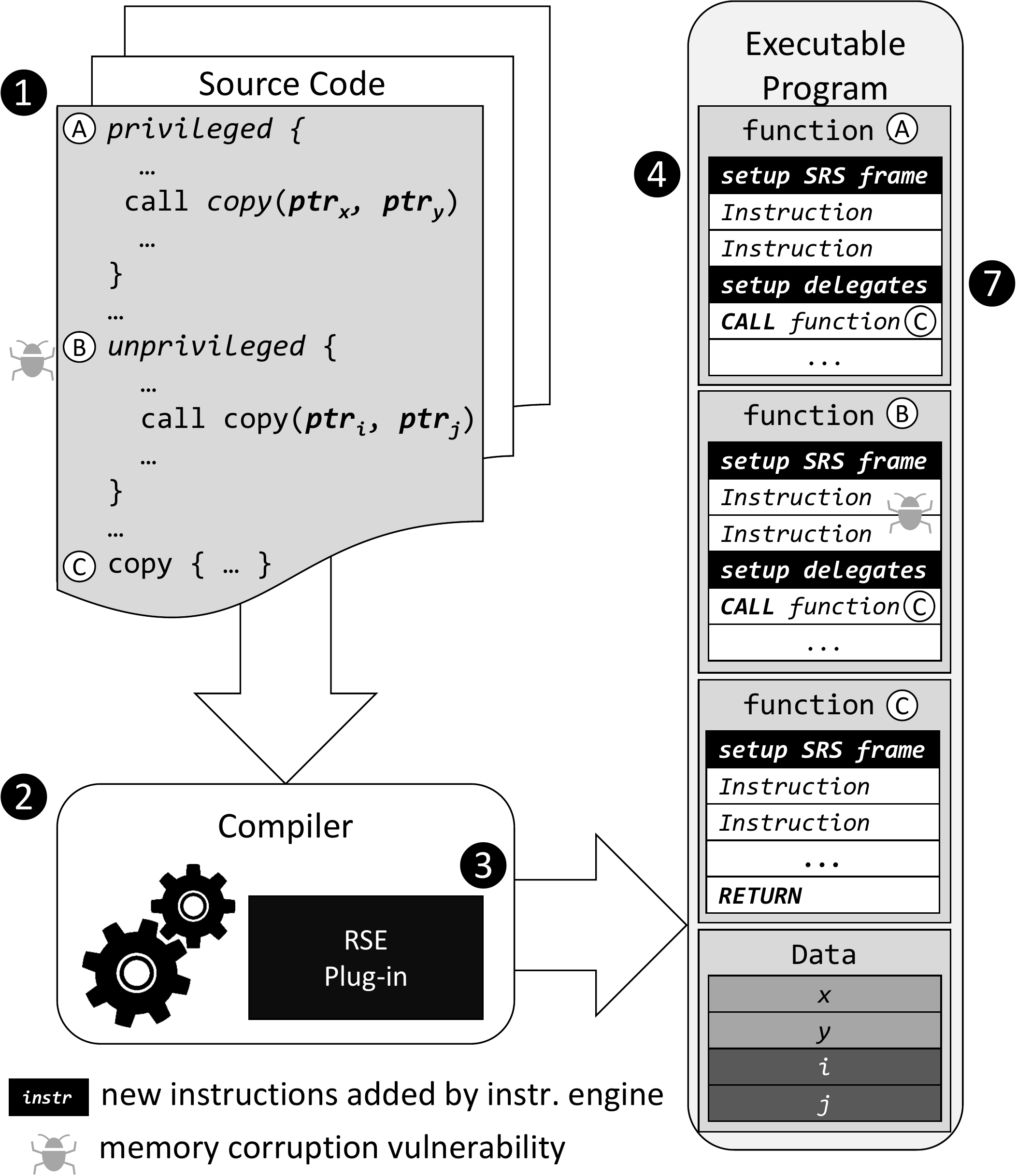}
\end{center}
  \caption{Compile-phase design of \SHORTNAME. Run-time memory accesses via pointers $ptr_{x}$, $ptr_{y}$ are limited to variables $x$ and $y$, while $ptr_{i}$, $ptr_{j}$ are limited to $i$ and $j$.}
  \label{fig:design}
\end{minipage}\hfill\begin{minipage}{\columnwidth}
\begin{center}
  \centering
  \includegraphics[width=0.75\columnwidth]{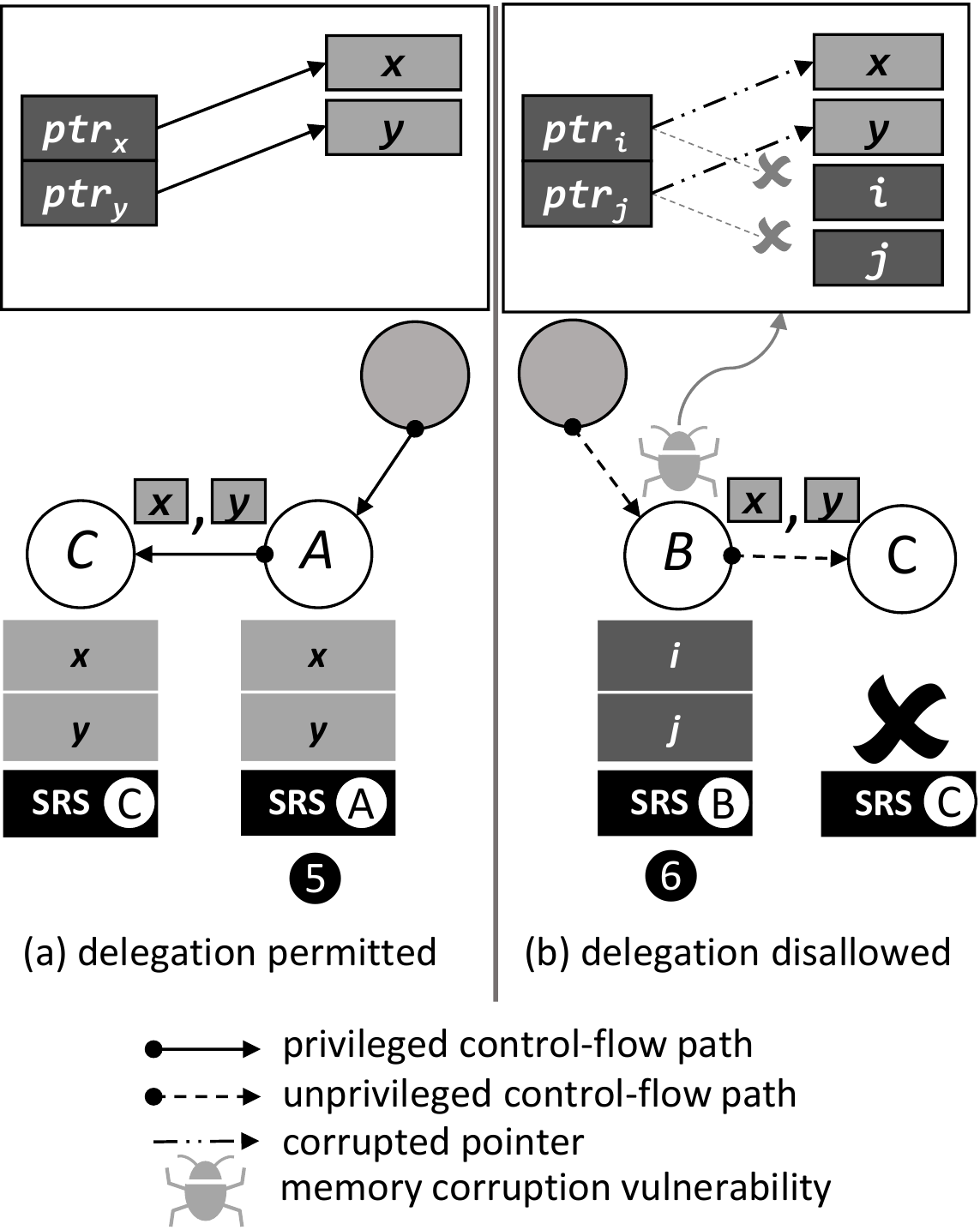}
  \caption{Run-time design of \SHORTNAME showing the call graph of program in Figure~\ref{fig:design}. In (a), access to variables $x$ and $y$ is successfully delegated from \texttt{A} to \texttt{C}. In (b), function \texttt{B} should not have access to $x$ and $y$ (e.g., $x$ could be a secret key), but a memory corruption vulnerability in \texttt{B} has been used to corrupt $ptr_{i}$ and $ptr_{j}$ to point to $x$ and $y$ instead of $i$ and $j$. \SHORTNAME prevents \texttt{B} from accessing or delegating $x$ and $y$.}
  \label{fig:call-graph}
\end{center}
\end{minipage}
\end{figure*}

Designing a solution to meet the requirements identified in Section~\ref{sec:requirements} requires addressing two major challenges:
\begin{enumerate}
\item \emph{Run-time enforcement:} enforcing variable scopes at run-time requires information which is usually only available at compile-time (necessary to meet \textbf{R1}).
\item \emph{Context-specific enforcement:} enforcing different rules for each invocation of a code block requires rules to be created, modified, and deleted dynamically at run-time (necessary to meet \textbf{R2}).
\end{enumerate}
\textbf{R3} implies that solutions to these challenges must be \emph{efficient}.
The high-level idea of \SHORTNAME is to extend the compiler to emit compile-time information about the visibility of variables, and to extend the underlying hardware to use this compiler-supplied information to dynamically create and update a set of memory access rules against which all memory accesses are checked. We chose function-level compartmentalization as the granularity of isolation, since this is sufficient to mitigate all currently known DOP attacks (Section~\ref{subsec:dop_mitigation}, Appendix~\ref{appx:otherdop}). However, \ABBRNAME can also be implemented at other granularities, without changes to the new \SHORTNAME hardware as described in Section~\ref{subsec:security_considerations} and Appendix~\ref{appx:ra-protection}.

\myparagraph{Run-time enforcement} 
Binary program code produced from languages such as C and C++ does not include information available to the compiler about variables and code blocks. \ABBRNAME needs this information to assign in-memory variables to specific \emph{execution contexts}. To bridge this gap between compile-time lexical scope and run-time execution context, we modified the compiler to instrument the program code with special instructions that record the variables that may be used by each block of code. \SHORTNAME introduces an instruction set extension for this purpose (Section~\ref{sec:instructions}). 

The compile-time components and behavior of \SHORTNAME are illustrated in Figure~\ref{fig:design}. 
An unmodified source code program (\dOne) is fed to the compiler (\dTwo), which checks that all variable accesses are correctly scoped (as usual). 
Our new \emph{\ABBRNAME Plug-in} (\dThree) in the compiler adds \SHORTNAME instructions (\dFour) at particular locations in the binary (e.g., at the start of functions). 
This results in a fully-functional program binary, instrumented with \SHORTNAME instructions.
These instructions are used by the \SHORTNAME hardware to create a set of rules against which all memory accesses can be checked at run-time.

\myparagraph{Context-specific enforcement}
Consider the program (\dOne) in Figure~\ref{fig:design}: function \texttt{C} receives two pointers as input and copies data from the first pointer to the second. 
It can be called from either function \texttt{A} or function \texttt{B} (the call graph is shown in Figure~\ref{fig:call-graph}).
In benign program execution, variables \textit{x} and \textit{y} are only used in a \emph{privileged} execution path, where access control checks prevent their misuse (e.g., $x$ could be a secret key). 
Function \texttt{B} contains an exploitable vulnerability allowing the attacker to control the pointers passed to function \texttt{C}. Since function \texttt{C} can be used to copy arbitrary data between two attacker-controlled pointers, this constitutes a usable DOP gadget. The attacker could use this to bypass the access control checks on variables \texttt{x} and \texttt{y} by accessing them through the unprivileged execution path. To enable context-specific enforcement, \SHORTNAME must be able to associate different memory access rules with \emph{each active instance} of a function.

To address this challenge, the \SHORTNAME hardware creates memory access rules dynamically for each individual function invocation, and stores these in a stack data structure called the \emph{Storage Region Stack} (SRS).
The SRS is kept in hardware-isolated \emph{protected memory}; only \SHORTNAME instructions can add or remove SRS entries.
Each entry in the SRS defines an area of data memory (e.g., the location of a variable) that may be accessed.
The SRS is organized into \emph{frames}; each frame corresponds to an execution context (i.e., contains all the entries for that context).
The topmost SRS frame corresponds to the active execution context.
On each memory access, e.g., load or store, \SHORTNAME validates that the memory address matches an entry in the topmost SRS frame.

To prevent the attack scenario in Figure~\ref{fig:call-graph}, \SHORTNAME's enforcement of function \texttt{C}'s memory accesses must distinguish between legitimate accesses to variables \texttt{x} and \texttt{y} when invoked by function \texttt{A}, and illegitimate accesses to them when invoked by the exploited function \texttt{B}. By default, \SHORTNAME prevents function \texttt{C} from accessing both \texttt{x} and \texttt{y}.

The SRS for function \texttt{A} (\dFive) includes variables \textit{x} and \textit{y}, and the SRS for function \texttt{B} (\dSix) includes variables \textit{i} and \textit{j} (Figure~\ref{fig:call-graph}).
To allow function \texttt{C} to access certain variables, the calling function must use a special instruction (Figure~\ref{fig:design} \dSeven) to \emph{delegate} access to a variable to function \texttt{C}: e.g., function \texttt{A} must delegate access to \textit{x} and \textit{y}. 
For valid delegation, the calling function must already have access to the delegated variables.
\SHORTNAME \ABBRNAME therefore prevents the attack shown in Figure~\ref{fig:call-graph}: 
even though the attacker can manipulate the pointers in function \texttt{B}, this function does not have access to \textit{x} and \textit{y} (its SRS lacks the corresponding entries) and hence function \texttt{B} cannot delegate access to these variables to function \texttt{C}.

\section{Implementation}
\label{sec:implementation}

To realize \emph{\LONGNAME} (\ABBRNAME), the new instructions must provide the following functionality:
\begin{enumerate}  \item Specify what storage regions are accessible by an execution context.
  \item Allow an execution context to dynamically delegate access to a storage region to another execution context (e.g., during function invocation or return).
  \item Subdivide a storage region so that partial access can be delegated.
\end{enumerate}
Section~\ref{sec:instructions} describes the \SHORTNAME instructions in detail.

Further, the compiler needs to be modified to emit these instructions to describe visibility rules derived from the program source code. Section~\ref{sec:instrumentation} describes our compiler \ABBRNAME Plug-in.

Finally, the underlying processor hardware must be extended to efficiently store and enforce the visibility rules described by the new instructions. Section~\ref{sec:hardware-impl} describes our processor modifications and proof-of-concept hardware realization of \SHORTNAME.

\subsection{\SHORTNAMETITLE Instructions}
\label{sec:instructions}

\SHORTNAME extends the RISC-V instruction set with six new SRS management instructions, as shown in Table~\ref{tbl:instructions}. 

The \texttt{sbent} and \texttt{sbxit} instructions are used by the compiler \ABBRNAME Plug-in to mark the beginning and end of each execution context. \SHORTNAME hardware uses these instructions to track when \SHORTNAME is first enabled and when execution context changes, and thus when new enforcement rules should be loaded in the SRS. When \texttt{sbent} is executed, a new frame is pushed on top of the SRS. Conversely, \texttt{sbxit} pops the topmost SRS frame. Program execution starts with an empty SRS and \SHORTNAME enforcement is initially disabled. The first function that supports \SHORTNAME, typically the program's \texttt{main} function, executes \texttt{sbent} to enable \SHORTNAME. \SHORTNAME remains enabled until a matching \texttt{sbxit} empties the stack. Due to the associated SRS management (explained in Section~\ref{sec:hardware-impl}), \texttt{sbent} and \texttt{sbxit} may consume up to $N$ additional cycles, where $N$ is the number of storage region entries in the topmost SRS frame. However, this only stalls the processor if the instruction is followed by another \texttt{sbent} or \texttt{sbxit} within the next $N$ executed instructions.

The \texttt{sradd} and \texttt{srdda} instructions create a storage region entry (\emph{SRS entry}) in the current (topmost) SRS frame. \SHORTNAME hardware uses these instructions to determine the bounds of memory areas that the current execution context is allowed to access. The value of the first register operand of each instruction sets the base address, and the second operand sets the limit address. An optional offset is added to to either the limit (\texttt{sradd}) or base (\texttt{srdda}) register operand.

The \texttt{srdlg} and \texttt{srdsub} instructions delegate an SRS entry from the currently executing function either to an invoked callee function or to the caller when the current function returns. \SHORTNAME hardware uses these instructions to derive SRS entries for flow of data which cannot be fully tracked by the compiler, such as context-specific accesses (Section~\ref{sec:design}). The \texttt{srdlg} instruction takes a single register operand and an immediate offset or only an immediate operand specifying an absolute address to determine which memory address to delegate. The resulting memory address is compared with the entries in the current SRS frame and if a match is found, the matching entry is copied to the \emph{next execution context entered}. If the delegation is followed by a \texttt{sbent}, the delegated entry is added to the newly created SRS frame. If the delegation is followed by a \texttt{sbxit}, the delegated entry is added to the caller's SRS frame. If multiple matching entries exist, only the most recently added entry is delegated.

The \texttt{srdsub} instruction is used to delegate a new SRS entry that is a subset of an existing SRS entry. It takes the same operands as \texttt{sradd}. \SHORTNAME hardware uses the operands to decide when storage region entries should be subdivided. If the new subdivided memory region is a subset of an existing SRS entry in the current SRS frame, a new SRS entry is created for a \emph{sub-region} using the new base and limit.

If no matching entry is found in the SRS when \texttt{srdlg} or \texttt{srdsub} execute, no entry is delegated. This prevents the use of \texttt{srdsub} to elevate the access rights of the next execution context beyond the rights of the current, but allows the delegation instructions to be applied to pointers which are not dereferenced directly in the current context. These include null-pointers and intentionally created out-of-scope pointers (e.g., via the use of pointer arithmetic) that are passed to callees for which they are in scope (e.g., accessor functions that receive opaque pointers from the caller). We refer to this as \emph{lax delegation} and describe a \emph{strict delegation} variant in Appendix~\ref{appx:strict_delegation}.    

The \texttt{srdlg} and \texttt{srdsub} instructions each consume one additional cycle if immediately followed by context switching \SHORTNAME instructions.

\begin{table}[t]
  \caption{\SHORTNAME Instructions. \normalfont
  \ifnotabridged \emph{Mnemonic} is used to refer to the instruction elsewhere in the paper. \fi 
  \ifnotabridged \emph{Name} is the full name of the instructions. \fi 
  \emph{Operands} lists the valid combinations of operands for each instruction: \texttt{r}$n$~is a register, \texttt{imm}~is an immediate operand, and \texttt{imm(r}\textit{n}\texttt{)}~is a register to which an optional immediate value is added as an offset.
  \emph{Cycles} indicates the effective number of cycles required at execute stage.}
  \label{tbl:instructions}
  \begin{center}
  \resizebox{\columnwidth}{!}{    \begin{tabular}{ l l l l }
      \hline
      \multicolumn{2}{l}{\small\textbf{Mnemonic \hspace{12pt} Name}} &
      \multicolumn{1}{c}{\small\textbf{Operands}} & 
      \multicolumn{1}{c}{\small\textbf{Cycles}} \\
      \hline

            \texttt{sbent}  & 
      \textbf{s}cope \textbf{b}lock \textbf{ent}er
      & n/a & 1 (+ $N$) \\

            \texttt{sbxit} &
      \textbf{s}cope \textbf{b}lock e\textbf{xit}
      & n/a & 1 (+ $N$) \\
      
            \texttt{sradd}  & 
      {\textbf{s}torage \textbf{r}egion \textbf{add}} & 
      \texttt{r1, imm(r2)} & 1 \\
      
            {\texttt{srdda}}  & 
      {\textbf{s}torage \textbf{r}egion \textbf{dda} (reverse add)} &
      \texttt{imm(r1), r2} & 1 \\

            \multirow{2}{*}{\texttt{srdlg}}  & 
      \multirow{2}{*}{\textbf{s}torage \textbf{r}egion \textbf{d}e\textbf{l}e\textbf{g}ate } & 
      \texttt{imm(r1)} & \multirow{2}{*}{1 (+ 1)} \\
      & & \texttt{imm} & \\

            {\texttt{srdsub}}  & 
      {\textbf{s}torage \textbf{r}egion \textbf{d}elegate \textbf{sub}-region} & 
      \texttt{r1, imm(r2)} & {1 (+ 1)} \\

      \hline
    \end{tabular}}
  \end{center}
\end{table}

\subsection{\ABBRNAME GCC Plug-in and Backend}
\label{sec:instrumentation}
\label{sec:compiler}

We developed an enhanced version of the GCC compiler incorporating a proof-of-concept \ABBRNAME GCC plug-in and a modified RISC-V backend that can automatically instrument C programs to benefit from \SHORTNAME without requiring any changes to program code or additional information from the developer (e.g., code annotations).

The GCC plug-in analyzes the program's \emph{Intermediate Representation} (IR) within GCC. The plug-in targets the high-level GIMPLE representation, which is a processor independent abstraction of the program. From the IR, the plug-in extracts information about the use of global and static variables in each function, the type of pointers passed as arguments in function calls, and the return type of each function to assess whether delegation is needed. The results of the analysis are passed to the modified RISC-V backend that operates on the low-level \emph{Register Transfer Language} (RTL) representation of the program and emits sequences of assembly. While the RTL lacks information about the lexical scope of variables, the backend supplements the information in the RTL with information retained from the prior \ABBRNAME plug-in analysis pass and emits \SHORTNAME instructions when expanding function prologues, epilogues and function call sites.

\myparagraph{Function instrumentation} Our modified RISC-V backend currently supports automatic instrumentation of C programs at \emph{function granularity} to protect the 1) the return address and other return state information, 2) stack variables, 3) arguments passed on the stack, 4) heap objects, and 5) global and static variables. The beginning of each distinct execution context is marked by inserting a single \texttt{sbent} instruction at the function call site just before the jump instruction. The end of an execution context is marked by inserting an \texttt{sbxit} instruction just before the return in the callee function. In Section~\ref{sec:evaluation} we show that function-level isolation is sufficient to mitigate all currently known DOP attacks. However, \ABBRNAME can also be implemented at other granularities, without changes to the new \SHORTNAME instructions or \SHORTNAME hardware as described in Section~\ref{subsec:security_considerations}.

\myparagraph{Return state and stack variables} By convention, the compiler adds a \emph{function prologue} to the beginning, and a \emph{function epilogue} to the end of each function. The function prologue is responsible for allocating space on the call stack for local variables, and storing the return address and old frame pointer to the stack. \SHORTNAME instructions are inserted before the standard prologue begins in order to create new SRS entries for local variables allocated by the prologue, as well as for any static or global variables accessed by the function.

Listing~\ref{lst:riscv-prologue} shows a function prologue (\dThree) that reserves space for a stack frame containing two 32-bit variables, the return address, and frame pointer (16~bytes in total) from the stack (line~8). It stores the value of the return address register (\texttt{ra}), and the register holding the frame pointer (\texttt{s0}) to the stack frame (lines~9, 10). Our instrumentation prepends the prologue with a \texttt{srdda} instruction (\dOne) that adds an SRS entry covering the whole stack frame (e.g., 16~bytes in Listing~\ref{lst:riscv-prologue}). \ifnotabridged The limit for this entry is one less than the current stack pointer value, and the base address is calculated by subtracting the size of the stack frame from the current stack pointer. \fi The compiler already knows the size of the stack frame since this is used to decrement the stack pointer (line~8). \ifnotabridged The SRS entry must be added before the standard prologue so that the prologue can access the stack to store the return address (line 9). \fi

\begin{lstlisting}[
  float,
  texcl,
  label=lst:riscv-prologue,
  caption={Function prologue instrumentation. Registers used are the stack pointer (\texttt{sp}), return address register (\texttt{ra}), frame pointer (\texttt{s0}), and temporary register zero (\texttt{t0}).}
]
; \dOne\ create SRS entry for local variables
srdda   -16(sp), sp ; base=sp-16, limit=sp
; \dTwo\ create SRS entry for a global variable
lui     t0,   addi    t0,t0,sradd   t0, 23(t0)
; \dThree\ standard prologue
addi    sp,sp,-16  ; decrement stack pointer
sw      ra,12(sp)  ; store ra at sp+12
sw      s0,8(sp)   ; store s0 at sp+8
addi    s0,sp,16   ; update frame pointer
\end{lstlisting}

\begin{lstlisting}[
  float,
  texcl,
  label=lst:riscv-memcpy,
  caption=Instrumented \texttt{memcpy()} call. \emph{dest} (\texttt{a0}) points to a buffer in the program's data section. \emph{src} pointer (\texttt{a1}) points to a local buffer allocated from the function's stack frame. Integer \emph{n} (\texttt{a2}) holds the number of bytes to copy.
]
; \dOne\ prepare arguments
lw      a0,-12(s0)  ; dest
add     a1,s0,-1060 ; src
li      a2,1024     ; n
; \dTwo\ setup delegations and callee srs context
srdlg   a0          ; delegate dest
srdsub  a1,1024(a1) ; delegate n bytes of src
sbent
; \dThree\ jump to \texttt{memcpy}
jal     ra,memcpy
\end{lstlisting}

\begin{lstlisting}[
  float,
  texcl,
  label=lst:riscv-funret,
  caption=Instrumented function returning a pointer via register \texttt{a0}.
]
; \dOne\ prepare return value
lw      a0,-12(s0)
; \dTwo\ delegate returned object and exit context
srdlg   a0  ; delegate returned object
sbxit
; \dThree\ return to caller
ret
\end{lstlisting}

\myparagraph{Global and static variables} In C programs, global objects can be accessed from any scope. However, \ABBRNAME instrumentation can effectively narrow the scope of global objects to only those functions that refer directly to these objects. For example, in Listing~\ref{lst:riscv-prologue}, a global \texttt{myobject} is accessed from the function's scope, so an SRS entry for \texttt{myobject} is created before the standard prologue begins (\dTwo). Separate SRS entries are added for each object accessed by the function. The sizes of global objects are known from the IR analysis and the address is evaluated by the linker. Static variables are handled similarly. Conversely, functions that access global objects indirectly (e.g., via function pointers) must receive the necessary SRS entries for these  objects via run-time delegation. Without such delegated rules \SHORTNAME would prevent access to indirectly accessed data.

\myparagraph{Function arguments and return values} Listing~\ref{lst:riscv-memcpy} shows an instrumented call to the \texttt{memcpy()} function in the C standard library. The  \texttt{memcpy()}  function  copies \emph{n} bytes from \emph{src} to \emph{dest}, so the caller prepares the arguments \emph{dest}, \emph{src}, and \emph{n} as usual (\dOne).  To allow \texttt{memcpy()} to operate on these memory areas, two delegation instructions are added to the program (\dTwo) just before the call (\dThree). The \emph{dest} pointer is held in register \texttt{a0} and points to a global buffer in the program's data section. The caller already has an SRS entry for this specific buffer, which it delegates using the \texttt{srdlg} instruction with register \texttt{a0} (line 6).

The \emph{src} buffer exists in the caller's own stack frame. 
To avoid giving \texttt{memcpy()} access to the caller's whole stack frame, the caller delegates only a \emph{sub-region} spanning only the \emph{src} buffer using \texttt{srdsub} (line 7). 
In this example, the programmer defined the number of bytes to copy (\emph{n}) based on the size of \emph{src}. If the size of \texttt{dest} is less than \emph{n}, this would result in a buffer overflow, which could be used to overwrite variables in the program's data section. However, since \texttt{memcpy()} is only delegated access to the memory area containing \texttt{dest}, \SHORTNAME prevents this memory error.

Listing~\ref{lst:riscv-funret} shows an instrumented function returning a pointer (\dOne). Before it returns (\dThree), the function delegates access to the returned object and exits the current execution context (\dTwo).

\myparagraph{Heap object allocation}
We implemented a wrapper on top of the C standard library \texttt{malloc()} function that creates SRS entries for the memory allocated on the heap, and delegates these to the caller.

\myparagraph{Deeply nested pointers} Keeping track of SRS entries is a challenge when deeply nested data structures are delegated, e.g., when the head of a linked list is passed as an argument. Our current PoC GCC Plug-in
cannot automatically infer the complete set of SRS entries that should be delegated when a callee receives a pointer to beginning of the nested pointer chain. Additionally, the number of SRS entries that must be accumulated and delegated for large nested data structures result in more frequent stalls at run-time (cf. additional profiling on CoreMark in Appendix~\ref{appx:coremark_config}).   

To handle delegation of such complex data structures, the \SHORTNAME GCC Plug-in must infer the relationship between linked data structures and emit instructions to allow \SHORTNAME hardware to derive the complete set of SRS entries that must be delegated for that structure. Our current GCC Plug-in implementation can derive SRS entries for structures allocated from memory pools known statically (a common pattern in embedded applications). However, to provide finer granularity policies that better match developer intent, a developer writing software for a \SHORTNAME-enabled platform can insert \texttt{srdlg} and \texttt{srdsub} instructions manually where needed. The \SHORTNAME support files include compiler macros for this purpose. Standard data structures can be provided with wrapper functions. Employing more sophisticated program analysis could insert such wrappers automatically.

\subsection{Hardware Implementation}
\label{sec:hardware-impl}

\begin{figure}[t]
  \centering
    \includegraphics[width=\hsize]{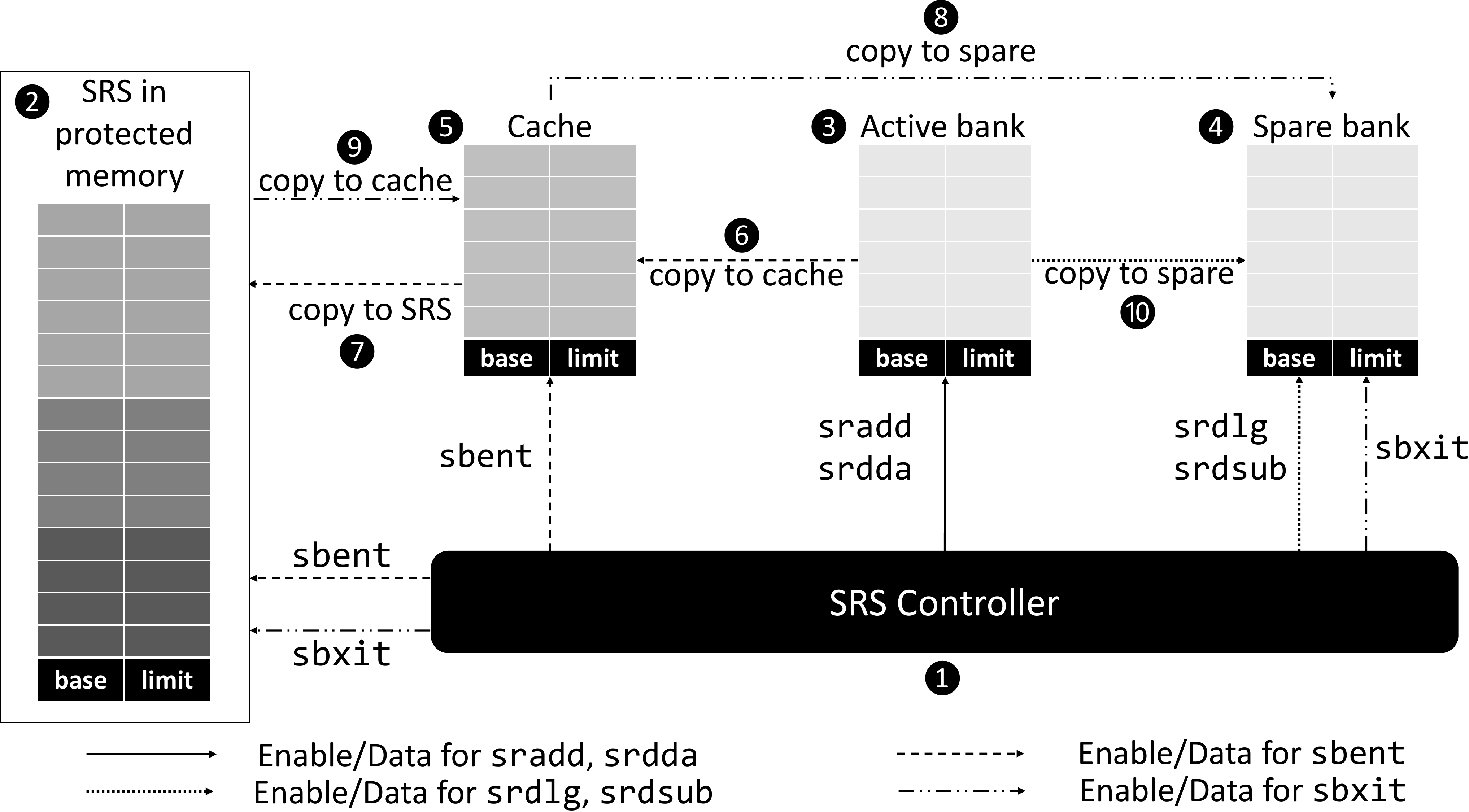}
    \caption{\SHORTNAME hardware architecture. On \texttt{sbent} and \texttt{sbxit} register banks \dThree\ and \dFour\ are exchanged.}
  \label{fig:impl}
\end{figure}

We developed a proof-of-concept hardware implementation of \SHORTNAME and extended it to the open-source RISC-V Pulpino core.\footnote{\url{http://www.pulp-platform.org/}}

We modified the instruction decoding stage of the processor pipeline to interpret the new instructions (Section~\ref{sec:instructions}). \ifnotabridged To minimize modifications to the decode stage all new instructions were encoded in RISC-V's existing S-type instruction format that allows up to two registers operands and a 12-bit signed integer immediate operand for each instruction. \fi After decoding, the appropriate control signals are sent to the \SHORTNAME unit, which realizes the execute stage of the new instructions.
Figure~\ref{fig:impl} shows the main components of the \SHORTNAME unit: the \emph{SRS controller} (\dOne), dedicated memory to hold the SRS (\dTwo), and three register banks (\dThree, \dFour, \dFive). The \emph{active bank} (\dThree) holds the entries in the SRS frame for the current execution context enabling each memory access to be compared against \emph{all active entries} efficiently.
The \emph{spare bank} (\dFour) holds entries delegated via \texttt{srdlg} and \texttt{srdsub} before a \SHORTNAME context switch occurs. It allows delegated entries for the \emph{next execution context} to be accumulated ahead of time. When a \SHORTNAME context switch occurs, the spare bank becomes the active bank (and vice versa), thus activating the delegated entries.
The third bank (\dFive) is used as a cache to hold a copy of the topmost frame of the SRS. This reduces the latency when the topmost SRS frame is transferred between stack memory and the spare bank.

When executing \texttt{sbent}, the controller activates the spare bank and transfers the contents of the currently active bank to the cache (\dSix) in a single cycle. The bank that held the previously active frame becomes the spare, and can be used for subsequent delegations. The entries in the cache must be stored for future use, and are transferred to the SRS in protected memory (\dSeven) over at most $N$ subsequent cycles, where $N$ is the maximum number of entries in the cache. During this time, the CPU continues to execute subsequent instructions normally until a new \SHORTNAME context switch occurs. Only if a \SHORTNAME context switch occurs before the cache has been emptied does the processor stall until the transfer is complete.

When executing \texttt{sbxit}, the controller copies the SRS frame from the cache into the spare bank (\dEight) while retaining  delegated entries (i.e., activating the entries that are already in the spare bank). The SRS frame in the previously active bank is no longer needed and is discarded. This executes in a single cycle. The cache, which now holds an out-of-date copy of the active frame, is updated with the topmost SRS frame from the protected memory (\dNine), which takes at most $N$ cycles, where $N$ is the number of entries in the topmost SRS frame in memory. This transfer to the cache does not stall the processor unless another \texttt{sbxit} is encountered before the cache is fully populated, in which case the CPU stalls until the next frame is available. However, if an \texttt{sbent} is encountered before the cache is fully populated, the partial cache is discarded and replaced with the contents of the active bank, without stalling.

The \texttt{sradd} and \texttt{srdda} instructions always operate on the active bank. When executing \texttt{srdsub}, the controller checks the active bank for an entry containing the given memory region and, if found, adds the new sub-entry to the spare bank. Similarly, in \texttt{srdlg}, the controller checks for the matching entry in the active bank and, if found, copies the entry to the spare bank (\dTen). The \texttt{srdlg} and \texttt{srdsub} instructions require an additional cycle only if followed immediately by a context switching \SHORTNAME instruction that modifies the spare bank.

Integrating \SHORTNAME into the processor pipeline at the execute stage also required modifying the memory access stage to intercept all memory access requests to the load/store unit. At each \texttt{load} or \texttt{store} instruction, the requested memory address and the number of requested bytes (one byte, half-word (two bytes), or word (four bytes)) are evaluated to a memory address range to be fetched from memory. The request is forwarded to the SRS controller, which compares it against all entries in the active bank. The registers in each bank are wired to comparators that enable all entries in the bank to be checked \emph{in parallel}.
If a match is found, i.e. the requested address range is a subset of any of the active entries, then the request is allowed to be evaluated by the processor's load/store unit, otherwise a hardware fault is raised. The memory access intercepted by \SHORTNAME executes without incurring additional cycles to \texttt{load} and \texttt{store} instructions.

\ifnotabridged

Since both \SHORTNAME instructions as well as \texttt{load} and \texttt{store} instructions require access to the \SHORTNAME hardware unit, we multiplexed shared access to it according to the currently decoded instruction. To ensure correct and hazard-free execution of the \SHORTNAME instructions within the pipeline, processor stalls due to \SHORTNAME instructions are engineered in the processor pipeline. \texttt{sbent} and \texttt{sbxit} instructions require that the processor stalls if another \texttt{sbent} or \texttt{sbxit} is encountered within $N$ cycles. The value $N$ corresponds to maximum number of entries in the cache bank. Hence, this is determined by the \SHORTNAME unit and indicated in stall control signals sent back to the decode stage which enforces the stall only when an \texttt{sbent} or \texttt{sbxit} instruction is encountered. The \texttt{srdlg} and \texttt{srdsub} instructions effectively need to stall for a an extra single cycle only if directly followed by context switching \SHORTNAME instructions. Being a pre-defined stall that only gets enforced if the next instruction is also a \SHORTNAME instruction, this is controlled directly in the decode stage and requires no feedback from the \SHORTNAME unit. Stall control for \SHORTNAME must also not conflict with other processor stalls associated with \texttt{load}, \texttt{store} and any \texttt{branch} instructions in case these directly follow or precede \SHORTNAME instructions. To also overcome potential data hazards, operand and data forwarding from preceding instructions is also supported for our \SHORTNAME instructions. This eliminates the need to inject additional stalls when results from preceding instructions are not yet updated in the registers but are required by the current \SHORTNAME instruction, which often occurs with \texttt{sradd} and \texttt{srdda} instructions that are preceded by auxiliary instructions that load immediate values to their operand registers.

\fi

We fully integrated \SHORTNAME with the Pulpino core and synthesized the extended processor on a Xilinx Zynq-7020 ZedBoard All Programmable SoC.\footnote{\url{http://zedboard.org}} The instruction-level functionality was validated by performing a register-transfer level (RTL) cycle-accurate simulation of the integrated hardware design using ModelSim/QuestaSim.\footnote{\url{https://www.mentor.com/products/fv/questa/}} We also extended \emph{Spike}\footnote{\url{https://github.com/riscv/riscv-isa-sim}}, the official RISC-V ISA simulator to support our \SHORTNAME instruction set extension.

\ifnotabridged

\subsection{Simulator Implementation}
\label{sec:emulator-impl}

We extended \emph{Spike}\footnote{\url{https://github.com/riscv/riscv-isa-sim}}, the official RISC-V ISA simulator to support our \SHORTNAME instruction set extension. Spike is part of the official RISC-V infrastructure and is currently the most accurate simulator for RISC-V assembler programs. It is regularly maintained by the RISC-V community, well integrated into the toolchain, and supports debugging with the GNU debugger (GDB).
We used it to analyze the security properties and performance profile of \ABBRNAME, and we have also made it available for developers and researchers who wish to reproduce our results or experiment with other uses of our \SHORTNAME instruction set extension.\footnote{The enhanced version of the Spike simulator is included with our accompanying materials at: \materials}

The simulator was extended by adding a SRS module to the Memory Management Unit (MMU) of the processor.
Similar to our hardware design (Section~\ref{sec:hardware-impl}), this module includes two banks of SRS registers for active and delegated entries, and a stack for inactive frames.
Each executed \SHORTNAME instruction is passed to our SRS module, which faithfully simulates the behavior of a real hardware implementation.
It also collects performance profiling statistics including the number of executed instructions, frequency of context switches, sizes of SRS frames, and the number of access checks performed.

\fi

\section{Evaluation}
\label{sec:evaluation}

To demonstrate the functionality of \SHORTNAME, we first show how it comprehensively mitigates Hu et al.'s~\cite{Hu16} DOP attack (described in Section~\ref{sec:background}), and in Appendix~\ref{appx:otherdop} we explain how it mitigates all other known DOP attacks, including the attack by Evans~\cite{Evans16}.
We then evaluate the security of \SHORTNAME with reference to the requirements defined in Section~\ref{subsec:requirements}, and analyze its performance and area overhead.

\subsection{DOP Mitigation Example}
\label{subsec:dop_mitigation}

We replicated the DOP attack by Hu et al.~\cite{Hu16} and ported the code to Pulpino to evaluate the effectiveness of \SHORTNAME. 
Although it was not possible to port the complete ProFTPD to our FPGA testbed or simulation environment, we concentrated our evaluation on the vulnerable \texttt{sreplace()} function.\footnote{For the same reason, it is also not possible to run standard SPEC benchmarks on this small microcontroller.}
We automatically instrumented this code with \SHORTNAME instructions using our GCC compiler extensions.
This means that all enforcement rules in the test programs are derived \emph{without any developer annotations} -- the GCC intermediate representation contains all information necessary for compile-time instrumentation, including: stack-frame sizes, global variable accesses, function calls, parameters, and return values.
We used our modified Spike simulator to trace the execution for both benign and malicious inputs, and verified that our instrumentation did not affect the correctness of the program under benign inputs.
The source code and instrumented symbolic assembler files are included in the supplementary material.\footnote{\materials}

We identified \emph{five} ways in which \ABBRNAME prevents this DOP attack. 
Any one of these would be sufficient to stop the attack, and thus the existence of five distinct mitigation points demonstrates the effectiveness of \ABBRNAME's layered defense strategy. 
All five were also verified experimentally using our modified Spike simulator.

\textbf{1)} The initial memory violation in \texttt{sreplace()} is caused by an out-of-bounds  \texttt{sstrncpy()} to a local stack buffer \texttt{buf}. The bound for the \texttt{sstrncpy()} call is calculated as \texttt{sizeof(buf) - strlen(pbuf)}. The contents of \texttt{pbuf} is attacker-controlled, and left without a trailing null terminator causes the subtraction to yield a negative value. Interpreted as an unsigned integer this causes \texttt{sstrncpy()} to overflow. \ABBRNAME instrumentation records and enforces the intended bounds of \texttt{buf} and \texttt{pbuf}, thus preventing the out-of-bounds access by \texttt{strlen()} and \texttt{sstrncpy()}.

\textbf{2)} The DOP program keeps internal state in unused areas of the program's data section. By default, \ABBRNAME denies access to such unused areas from all functions. The attacker could attempt to work around this by using pre-existing global variables. However, because access to globals can be narrowed by the use of \ABBRNAME, all the DOP gadgets must either share access to the same global data structure, or all be reachable by data flows to and from this data structure. Gadgets that legitimately share access to such data are more likely to use this data in benign program operation. This is undesirable for DOP because re-purposing such data could have unwanted side-effects on program execution, or be overwritten during benign operations, thus significantly limiting the amount of data or the set of gadgets that can be used in the attack. 

\textbf{3)} The exploit corrupts variables in global data structures to control the operands of the addition and dereference gadgets. During benign execution, the \texttt{sreplace()} function should only operate on a copy of the \texttt{main\_server.ServerName} pointer passed by value, and on unrelated fields in the global \texttt{session} structure, also passed by value. Therefore, \ABBRNAME denies  \texttt{sreplace()} access to these global variables, and thus blocks the addition and dereference gadgets by preventing the attacker from controlling their respective operands.

\textbf{4)} The dereference gadget accesses \texttt{ssl\_ctx} via a pointer included in the DOP payload and traverses the chain of linked structures to determine the location of the secret key. Since \texttt{ssl\_ctx} is defined as a global static in the \texttt{mod\_tls.c} source file, \ABBRNAME prevents the dereference gadget from accessing this structure or linked structures.

\textbf{5)} Once the address is known, \texttt{sreplace()} is used to corrupt a local static \texttt{mons} array containing string pointers. The \texttt{mons} array (Appendix~\ref{appx:pr_strtime}) contains pointers to string literals in the program's data section that are used by the \texttt{pr\_strtime()} function to format human readable dates. Each pointer in \texttt{mons} is redirected to the same memory location. One byte of the secret key is then copied to that location. Whenever any date is formatted by the corrupt \texttt{pr\_strtime()}, it leaks a few bytes of the key to the attacker. The process is repeated until the entire key has been extracted. \ABBRNAME prevents this exfiltration in two ways: Firstly, because the scope of \texttt{mons} is local to the \texttt{pr\_strtime()} function, \texttt{sreplace()} cannot overwrite it with new pointers. Secondly, \ABBRNAME ensures that \texttt{pr\_strtime()} cannot access the key, even if it attempts to dereference corrupt \texttt{mons} pointers.

Although \ABBRNAME cannot guarantee the absence of usable data-oriented gadgets in arbitrary programs, this example demonstrates how \ABBRNAME significantly limits the expressiveness of DOP attacks in programs with at least some degree of structural data separation, e.g., by minimizing the scope of variables (see Section~\ref{sec:assumptions}), as is the case in virtually all real-world programs.

\subsection{Security Considerations}
\label{subsec:security_considerations}

\myparagraph{R1. Multi-granularity enforcement} With the appropriate instrumentation, \SHORTNAME instructions can be used to enforce memory protection at run-time for any granularity of protection domain and any granularity of protected region.
Our enhanced GCC compiler automatically emits \SHORTNAME instructions at function-level granularity which, as shown in Section~\ref{sec:background}) and Appendix~\ref{appx:otherdop} is \emph{necessary} and \emph{sufficient} to thwart currently known DOP attacks. 
However, \SHORTNAME can also be used to enforce policies with either coarser or finer granularity of execution contexts. 
This is possible because \SHORTNAME instructions are agnostic to programming language and language constructs. 
We show an example of finer granularity in Appendix~\ref{appx:ra-protection} by isolating the function prologue and epilogue from the function body.
Thus even if the function body contains a memory error, this cannot be used to corrupt the return address on the stack (i.e., prevents control-flow hijacking).
For both of the above granularities, all necessary information about the protection domains can be deduced by the compiler, thus allowing automatic instrumentation.
Instrumentation at other granularities could also be inferred from existing language constructs (e.g., loops) or may require developer annotations.

\ifnotabridged
As another example, consider the simple loop in Listing~\ref{lst:c-loop}. 
The \texttt{for} loop forms a separate code block in terms of lexical scope. 
Since the index variable \texttt{i} is declared in the loop signature, it cannot be accessed from outside the \texttt{for} loop. 
The statement in the loop body accesses the \texttt{name} array, the \texttt{buf} array, and the \texttt{len} integer. 
The loop can be isolated in a separate execution context from the rest of the function body by surrounding it with \texttt{sbent} and \texttt{sbxit} instructions. 
Access to \texttt{name} and the buffer referenced by \texttt{buf} are delegated to the loop's execution context via \texttt{srdlg}, and access to the variable \texttt{i}, which is not accessed from outside the loop, is delegated with \texttt{srdlgm}. 
This ensures that the \texttt{for} loop is executed with minimal privileges. Should the value of \texttt{len} exceed the size of the \texttt{name} buffer, \ABBRNAME prevents the buffer operation from overflowing into the \texttt{password} array, which should only be accessed later in the function body.

In addition to execution contexts deduced from C control constructs, such as loops, the programmer may use \emph{unnamed blocks} to group related code and data together. 
Any variables declared inside an unnamed block are considered by the compiler to exist only within the block's lexical scope. 
\SHORTNAME can make use of this \emph{standard} language feature to automatically infer developer intent when determining execution contexts during instrumentation.

\begin{lstlisting}[
  float,
  style=customc,
  label=lst:c-loop,
  caption=C program with a \texttt{for} loop containing a potentially vulnerable memory error. There is no bounds checking when writing to the \texttt{name} buffer.
]
static char name[40];
static char password[16];

void fun(char *buf, int len) {
  int x = len - 1;

  for (int i = 0; i<len; i++, x--) {
    name[i] = buf[x];  // overflows when len > 40
  }

  // do something with password
}
\end{lstlisting}
\fi

\myparagraph{R2. Context-specific enforcement} In \SHORTNAME, the set of active SRS entries can differ between different invocations of the same subject, depending on which entries have been delegated to this subject (e.g., variables passed to a function by its caller or callee functions).

\myparagraph{R3. Complete mediation} Since \SHORTNAME hardware checks every memory access against the currently active set of SRS entries, a memory access without a matching entry will fail.
Therefore, the only possible memory accesses are those that would be allowed by a compile-time check.
We discuss the scalability, performance and area overhead in Section~\ref{subsec:performance_evaluation}.

\myparagraph{Preventing confused deputies} In a \emph{confused deputy attack}, the attacker attempts to subvert the \ABBRNAME property by misusing existing \SHORTNAME instructions at run-time to create unintended rules (i.e., rule-creating instructions are the confused deputies).
Our design ensures that no such instructions are available to the attacker.
Instructions that create rules for static allocations (stack and global variables) are encoded directly into the instrumentation.
Since these cannot be modified at run-time, they cannot be used as confused deputies.
Instructions that create rules for dynamic allocations could potentially be used as confused deputies, but this is practically infeasible because these instructions are only found within memory allocators e.g., malloc().
It is reasonable to assume that memory allocators are trusted (or at least that the absence of run-time vulnerabilities can be easily verified). We recommend that manually annotated code is vetted for allocators that create arbitrary rules at run-time.
Furthermore, an attacker can only initiate a confused deputy attack if he already controls some part of the code, which is very difficult since every memory access in the instrumented program is checked by the \SHORTNAME hardware.

\myparagraph{Interfacing with legacy code}
Legacy code, such as pre-compiled shared libraries, can be instrumented using wrapper functions. 
For example, our \texttt{malloc()} wrapper (Section~\ref{sec:instrumentation}) provides function-granularity isolation for delegated objects and the stack frame, and provides coarse-grained module-level isolation for other library data.

\myparagraph{Mitigating DOP attacks} As shown above, \SHORTNAME fulfils all requirements defined in Section~\ref{subsec:requirements}, and as shown in Section~\ref{subsec:dop_mitigation} and Appendix~\ref{appx:otherdop}, function-level granularity is sufficient to mitigate all currently known DOP attacks (at multiple points in each attack).

\myparagraph{Mitigating ROP attacks} Additionally, \SHORTNAME can defend against ROP attacks.
As explained in Appendix~\ref{appx:ra-protection}, the function prologue and epilogue are placed in a separate execution context from the function's main body.
By restricting a function's return state information only to the prologue/epilogue's execution context, \SHORTNAME protects this information from the potentially corrupted main body of the function.
Without the ability to control the function's return value, an attacker cannot mount ROP attacks.

\subsection{Performance and Area Overhead}
\label{subsec:performance_evaluation}

We evaluated the performance and area overhead of \SHORTNAME using the extended Pulpino processor synthesized using Xilinx Vivado\footnote{\url{https://www.xilinx.com/products/design-tools/vivado.html}} for the ZedBoard 7020 prototyping kit.\footnote{\url{http://zedboard.org}} \ifnotabridged
For the performance evaluations, we used both the ProFTPD code excerpt (Section~\ref{subsec:dop_mitigation}) as well as the CoreMark processor benchmark.\footnote{\url{http://www.eembc.org/coremark/index.php}}

\myparagraph{Performance evaluation (ProFTPD excerpt)}
This program consists of 527~lines of C code and its deepest call chain is seven levels deep.
Table~\ref{tbl:proftpd_overhead_cycles} shows the total cycle count for each type of \SHORTNAME instruction, and the number of cycles for which the processor was stalled when running on a \SHORTNAME equipped core, compared with the unmodified program.
In the instrumented version, the cycle counts for already existing instructions, including load/store, are not affected.
Since rules are checked in parallel, the number of enforcement rules per subject does not impact performance up to the number of available \SHORTNAME registers.
Taking into account the added \SHORTNAME instructions, the processor stalls, and the additional instructions needed to support the instrumentation, the total performance overhead was 1.6\%.

\fi

\myparagraph{Performance evaluation (CoreMark)}
CoreMark is a synthetic CPU performance benchmark for embedded systems based on a realistic mixture of commonly used algorithms including matrix and linked list manipulation, state machine operations, and Cyclic Redundancy Checks (CRCs).\footnote{\url{http://www.eembc.org/coremark/faq.php}}
It consists of \textasciitilde1500~lines of C code, with a deepest call chain of 11 levels.
All instrumentation in CoreMark was automatically generated by our extended GCC compiler, and we ran the benchmark on the Pulpino SoC extended with \SHORTNAME instructions. Binary size increased by 11\% as a result of instrumentation.
Table~\ref{tbl:coremark_overhead_cycles} shows the total number of executed \SHORTNAME instructions, the number of consumed cycles, and the number of cycles the processor was stalled for a single iteration of CoreMark, compared with an unmodified version. The added instructions account for a performance overhead of 3.1\%. We also ran CoreMark with varying iteration counts on the FPGA and observed an average overall performance overhead of 3.2\%.
The number of entries per SRS frame varied between 1 and 23, with a maximum of 11 frames.

 \ifnotabridged
\begin{table}[t]
  \caption{Breakdown of \SHORTNAME performance overhead by instruction for ProFTPD excerpt.}
  \label{tbl:proftpd_overhead_cycles}
  \begin{center}
    \resizebox{\columnwidth}{!}{    \begin{tabular}{ l | r r r | r r }
      & \textbf{\# instr} & \textbf{\# cycles} & \textbf{\# stalls} & \textbf{overhead}$^{\mathrm{a}}$\\
      \hline
              \texttt{sbent}  & 759 & 994 & 235 & 0.4\%\\
        \texttt{sbxit}  & 759 & 807 &  48 & 0.3\%\\
        \texttt{sradd}  & 348 & 348 & -   & 0.1\%\\
        \texttt{sradd}  & 759 & 759 & -   & 0.3\%\\
        \texttt{srdlg}  & 819 & 819 & -   & 0.3\%\\
        \texttt{srdsub} & 23  &  23 & -   & 0.0\%\\
        \textit{Other}  & 641 & 641 & -   & 0.2\%\\
        \hline
        \textbf{\SHORTNAME total} & \textbf{4108} & \textbf{4391} & \textbf{283} & \textbf{1.6\%} \\
      \hline
      \multicolumn{5}{p{\mytablefootnotewidth}}{\footnotesize\raggedright$^{\mathrm{a}}$Calculations based on uninstrumented ProFTPD~excerpt (276463~cycles).} \\
    \end{tabular}}
  \end{center}
\end{table}
\fi

\begin{table}[t]
  \caption{Breakdown of \SHORTNAME performance overhead by instruction for one iteration of CoreMark.} 
  \label{tbl:coremark_overhead_cycles}
  \begin{center}
    \resizebox{\columnwidth}{!}{    \begin{tabular}{ l | r r r | r r }
      & \textbf{\# instr} & \textbf{\# cycles} & \textbf{\# stalls} & \textbf{overhead}$^{\mathrm{a}}$\\
      \hline
              \texttt{sbent}  & 2512 &  3184 & 672 & 0.7\%\\
        \texttt{sbxit}  & 2512 &  3083 & 571 & 0.7\%\\
        \texttt{sradd}  & 318  &   318 & -   & 0.1\%\\
        \texttt{sradd}  & 2511 &  2511 & -   & 0.5\%\\
        \texttt{srdlg}  & 1379 &  1459 & 80  & 0.3\%\\
        \texttt{srdsub} & 654  &   990 & 336 & 0.2\%\\
        \textit{Other}  & 2742 &  2964 & -   & 0.6\%\\
        \hline
        \textbf{\SHORTNAME total} & \textbf{9886} & \textbf{11544} & \textbf{1243} & \textbf{3.1\%} \\
      \hline
      \multicolumn{5}{p{\mytablefootnotewidth}}{\footnotesize\raggedright$^{\mathrm{a}}$Calculations based on uninstrumented CoreMark (458150~cycles).} \\
      \multicolumn{5}{p{\mytablefootnotewidth}}{\footnotesize\raggedright\hspace{5pt}CoreMark configuration details are in Appendix~\ref{appx:coremark_config}.} \\
    \end{tabular}}
  \end{center}
\end{table}

\begin{figure}[t]
  \centering
    \includegraphics[width=\hsize]{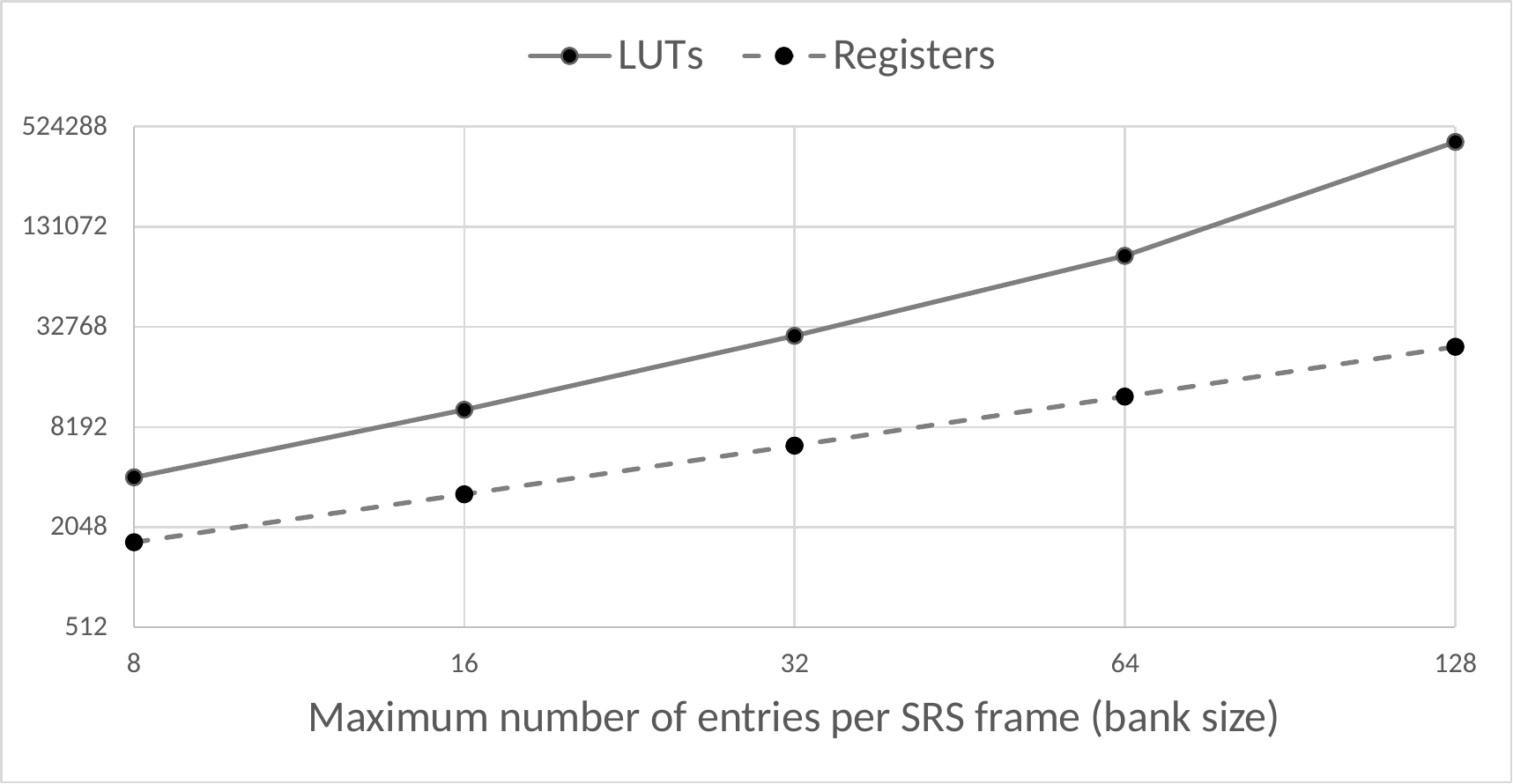}
    \caption{Logic resources (LUTs and registers) required to support different numbers of entries per SRS frame and protected memory for $16$ frames.}
  \label{fig:areaplot}
\end{figure}

\myparagraph{Area and memory overhead} We synthesized \SHORTNAME using Xilinx Vivado for a Xilinx Zynq-7020 ZedBoard (Virtex-7 XC7Z020 FPGA).
Figure~\ref{fig:areaplot} shows the number of look-up tables (LUTs) and registers required to support different numbers of entries per SRS frame.
As expected, the area overhead increases linearly with the bank size (i.e., the maximum number of entries per frame), since more entries must be checked in parallel.

To support a bank size of $16$ and a maximum of $16$ frames (i.e., a protected memory size of $16 \times 16$ entries), \SHORTNAME uses $10,376$ LUTs, $3,221$ registers, and one 18~kB block RAM.
This is less area than a 128-bit high-throughput pipelined hardware AES cipher, which uses $12,475$ LUTs and $10,769$ registers.\footnote{\url{https://opencores.org/project,aes-128_pipelined_encryption}} The Pulpino SoC itself uses $15,444$ LUTs and $9,758$ registers.
We also synthesized this \SHORTNAME configuration using Synopsys Design Compiler targeting the NanGate 45~nm Open Cell Library\footnote{http://www.nangate.com/} which gave a logic size of approximately 800,000 transistors.
In comparison to a modern general-purpose SoC, such as the Apple A10 quad-core ARM64 mobile SoC (3.3~billion transistors), the area overhead of \SHORTNAME is negligible.
Thus, \SHORTNAME is suitable for deployment on a wide range of SoCs, including small MCUs like the Pulpino.

\section{Related Work}
\label{sec:related-work}

\begin{figure}[t]
  \centering
    \ifanonymous
    \includegraphics[width=\hsize]{figures/related-work-graph-anon}
    \else
    \includegraphics[width=\hsize]{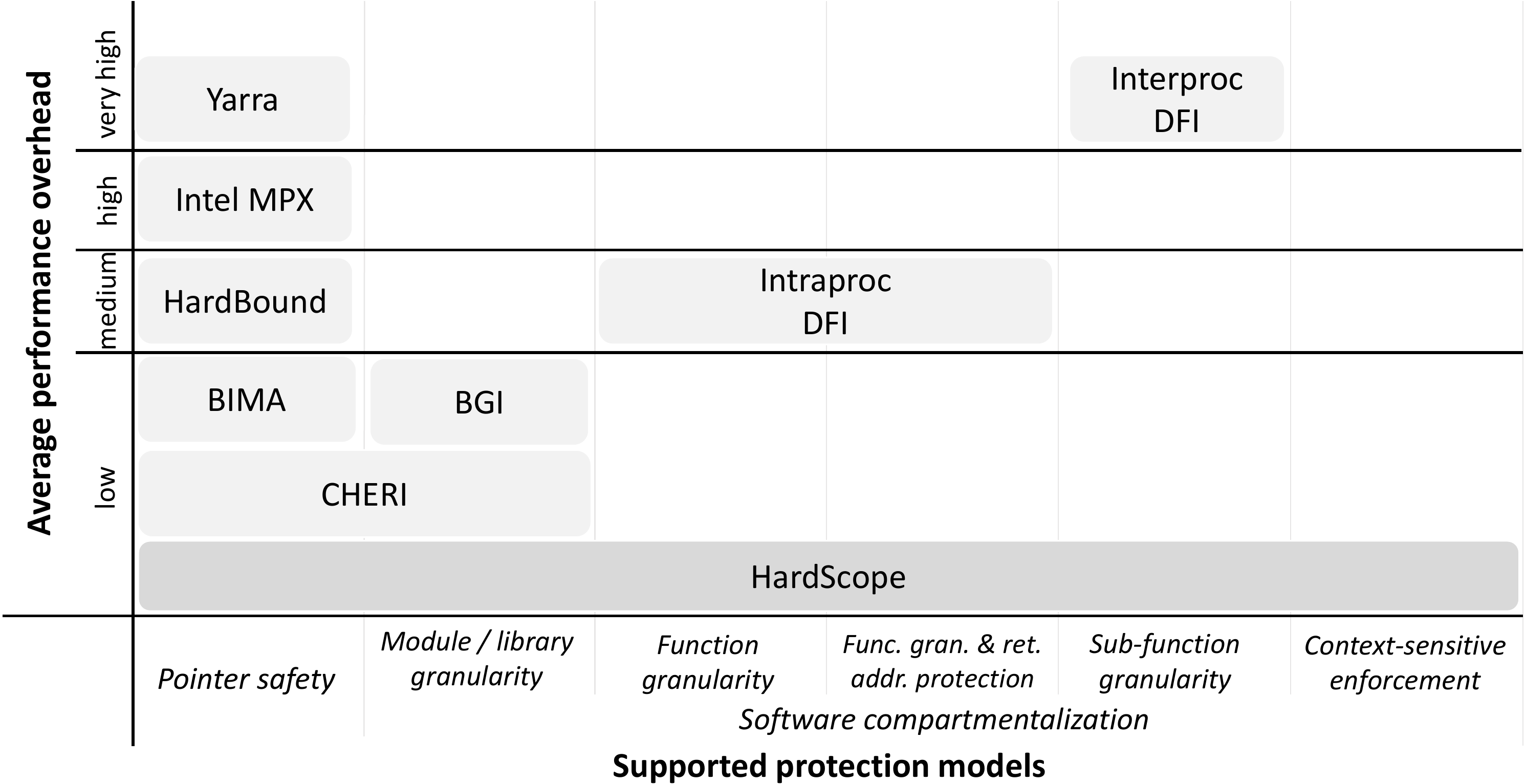}
    \fi
    \caption{Taxonomy of defenses by security model and overhead. The figure shows only defenses that can support a variable number of security domains.}
  \label{fig:taxonomy}
\end{figure}

Various software-only and hardware-assisted memory safety technologies have been proposed and/or deployed (e.g., \cite{Abadi09,Akritidis08,Bhatkar08,Cadar08,Castro06,Castro09,Devietti08,Erlingsson06,Kuvaiskii17,Kuznetsov14,Kwon13,Schlesinger11,Serebryany12}).
We discuss those that aim to mitigate non-control-data attacks. Figure~\ref{fig:taxonomy} shows a taxonomy of defenses that can instantiate policies effective against DOP. To the best of our knowledge, \ABBRNAME is the first scheme that specifically considers DOP attacks in its threat model.

\emph{\bf Pointer Safety} can prevent \emph{spatial memory errors} that are exploited in memory corruption attacks. Typical realizations associate a base address and upper bound with each pointer, and check that the memory accesses that occur when dereferencing the pointer fall within those bounds. We call these schemes \emph{pointer-oriented}. The associated bounds metadata can either be stored with each pointer or in a disjoint area of memory. 

\emph{Fat-pointer} schemes add bounds metadata directly to the pointer e.g., by increasing the length of the pointer~\cite{Necula02} or borrowing unused bits from the pointer~\cite{Kuvaiskii17}. This incurs only a small memory overhead but changes the memory layout of program in ways that break both binary and source code compatibility. 

BIMA~\cite{Kwon13} is a hardware-assisted fat-pointer scheme developed within the SAFE project.\footnote{\url{http://www.crash-safe.org/}} BIMA encodes pointer bounds metadata within the pointer itself (e.g., on a 64~bit system, it assumes that 46~bit addresses are sufficient).
On the \emph{simplified SAFE processor}, a clean-slate ISA design which includes various security enhancements, this scheme has no performance overhead and worst-case 16\% memory overhead.
However, the compact encoding of pointer metadata results in alignment restrictions on pointers. This necessitates the use of custom stack allocators when applying this approach to stack data structures~\cite{Duck17}.

\emph{SoftBound}~\cite{Nagarakatte09} and \emph{HardBound}~\cite{Devietti08} use pointer bounds metadata stored in disjoint \emph{shadow space} memory to ensure pointer safety. Although this retains the program's original memory layout, it requires bounds information to be fetched from the shadow space before checks. SoftBound breaks cache locality and leads to additional cache misses when retrieving pointer bounds, incurring an average performance overhead of 67\% in standard benchmarks. HardBound is a hardware-assisted scheme where the processor checks associated pointer bounds implicitly when a pointer is dereferenced. HardBound incurs an average performance overhead of $\approx$10\%. Both schemes exhibit a worst-case memory overhead of $\approx$200\%.

Intel's Memory Protection Extensions\footnote{\url{https://software.intel.com/en-us/isa-extensions/intel-mpx}} (MPX), introduced in the Intel Skylake microarchitecture in late 2015, is currently the only example of hardware-assisted pointer safety technology being deployed in real systems. MPX adds new instructions for performing bounds checks on pointers. MPX stores bounds metadata in a disjoint memory area. Bounds information for up to four pointers can be stored in dedicated registers for fast checks. Oleksenko et al.~\cite{Oleksenko17} found that MPX incurs an average performance overhead of 50\% and a memory overhead of 1.9x, largely due to the time and memory required for storing and loading bounds metadata.

Unlike the above pointer-oriented schemes, RSE (e.g., \SHORTNAME) is \emph{block-oriented} as it associates access control rules with blocks of code, rather than individual pointers. However, as \SHORTNAME policies can be applied at the granularity of even single instructions, it can also enforce pointer safety. \SHORTNAME retains the program's original memory layout and does not require special alignment of pointers. In addition, \SHORTNAME also validates that pointer dereferences occur from legitimate execution contexts.

\ifnotabridged
\emph{Red-Zone Tripwires} can be used to ensure partial pointer safety against contiguous overflows. By placing a block of invalid memory that acts as a "red-zone" between memory objects. Loads and stores are instrumented to verify if the red-zone is tripped. Contiguous overflows, e.g. past an array boundary will hit the red-zone tripwire. This provides only partial pointer safety, as non-contiguous accesses or accesses with a larger step distance than the size of the red-zone can violate spatial safety without setting of the tripwire Modern compilers, such as GCC and LLVM, support instrumenting code operating with red-zone tripwire based run-time bounds checks via AddressSanitizer~\cite{Serebryany12}. 
\fi

\emph{\bf Pointer authenticity} can ensure the unforgeability of pointers, preventing non-control-data attacks and use of DOP gadgets that rely on retargeting pointers.

\emph{Yarra}~\cite{Schlesinger11} is  a variant C that ensures the authenticity of a pointer's type for \emph{critical data types} ascribed by the programmer. YARRA guarantees that such critical data is only written through pointers with the given static type. However, YARRA incurs a prohibitively high overhead when used for whole program protection (4x - 6x). 

\emph{PointGuard}~\cite{Cowan03} instruments programs to encrypt all pointers at run-time by XORing them against a key generated at program initialization. Pointers are decrypted before dereference. PointGuard incurs a small to medium overhead (0\% - 20\%) in real-world programs, but is vulnerable to information disclosure e.g., if the ciphertext of a known pointer becomes known to an attacker.

\emph{ARMv8.3 Pointer Authentication}~\cite{PAC} is a hardware-assisted mechanism in the ARMv8.3 processor architecture that ensures the authenticity of pointers by calculating a \emph{Pointer Authentication Code} (PAC) as a keyed MAC of the pointer value and a 64-bit context (e.g., the current value of the stack pointer). The PAC is stored in the unused bits of 64-bit pointers and verified before dereferencing the pointer to ensure its authenticity. The inclusion of the context value prevents unauthorized copying of the pointer and its PAC to another context within the program. The ARMv8.3 architecture provides four keys for PAC (two for code pointer / two for data pointers) and
a fifth key usable for general purpose authentication code generation. The keys are stored in internal CPU register and are not accessible from user mode, but must be managed by privileged software (e.g., ephemeral keys per process for user mode, or per boot for kernel mode). 

\SHORTNAME does not provide pointer authenticity, but when applied at a fine granularity, it can greatly reduce the attack surface of pointers and non-pointer data. 

\emph{\bf Software compartmentalization} aims to mitigate the consequences of memory vulnerabilities by isolating software components into distinct protection domains. 

\emph{Software Fault Isolation} (SFI)~\cite{Wahbe93a,Erlingsson06,Mao11} compartmentalizes software at a module level e.g., kernel modules and dynamically loaded libraries. Non-control-data and DOP attacks cannot interact with data outside the module boundary but attacks that operate fully within the confines of a single module remain viable. \emph{Byte Granularity Isolation} (BGI)~\cite{Castro09} is an SFI variant that instruments kernel extensions and can enforce access control policies at fine data granularity with moderate overhead (0\%-16\%).

\emph{Data-Flow Integrity} (DFI)~\cite{Castro06} is a software-only approach for mitigating control-flow and non-control-data attacks. At compile-time, static analysis constructs a data-flow graph of a program. The code is instrumented to record the last instruction that wrote to each variable. On every read, the origin of the last write is checked against the pre-computed data-flow graph. Like \SHORTNAME, DFI can be instantiated at various granularities. \emph{Intraproc DFI} only instruments uses of control-data and uses of local variables without definitions outside their function, thus providing function-granularity isolation for stack data. \emph{Interproc DFI} isolates individual data flows from each other. Intraproc DFI incurs 46\% and Interproc DFI incurs 104\% performance overhead and 50\% memory overhead. Hardware-Assisted Data-flow Isolation (HDFI)~\cite{Song16} provides instruction-level granularity isolation by tagging each machine word in memory and every memory access instruction with a protection domain. However, it only supports two simultaneous protection domains. 

\emph{Probabilistic schemes} aim to randomize the data or its layout at run-time so that unauthorized accesses would have unpredictable results.
Data Randomization~\cite{Cadar08} uses static analysis to partition code into equivalence classes, and then instruments all load/store operations to XOR the data with a class-specific mask.
Data Space Randomization~\cite{Bhatkar08} randomizes the layout of data in memory.
However, probabilistic schemes rely on some secret information (e.g., the XOR mask or randomization secret) which if leaked or inferred by the attacker could undermine the scheme.

\emph{Hardware-architectures} that enable different protection models have been proposed. Fine-grained tagged memory systems e.g., \emph{lowRISC}\footnote{\url{http://www.lowrisc.org/downloads/lowRISC-memo-2014-001.pdf}} can be used to assist the implementation of sophisticated memory access policies, including RSE. However, unlike \SHORTNAME, lowRISC only differentiates between access types (read/write) and can not apply different policies per subject without reprogramming. \emph{Intel Memory Protection Keys} (MPK)\footnote{\url{https://www.kernel.org/doc//x86/protection-keys.txt}} provides hardware support to associate memory at page granularity to one of 16 distinct protection domains, but unlike \SHORTNAME, it does not support context-specific policies or delegation.

\emph{CHERI}~\cite{Woodruff14} is hardware-assisted capability model that extends the 64-bit MIPS ISA with byte-granularity enforcement of memory accesses. CHERI can support various protection models, such as pointer safety~\cite{Woodruff14} and software compartmentalization~\cite{Watson15} at library or module level. However, programs must be re-engineered by hand to benefit from CHERI. 

Run-time attestation, such as \emph{Control-Flow Attestation}~\cite{Abera16, Dessouky17} can detect, but not prevent, both control-flow and some non-control-data attacks.

Although \SHORTNAME shares many of the same goals as the above schemes, it differs in several fundamental aspects. Compared to software-based schemes (e.g., DFI~\cite{Castro06} and SoftBound~\cite{Nagarakatte09}), \SHORTNAME has significantly lower overhead, does not require whole-program static analysis, and can enforce different rules during different invocations of the same function (context-specific memory isolation). \SHORTNAME RSE policies can be instantiated for a large class of programs without additional input from developers (cf., YARRA~\cite{Schlesinger11}), or software re-engineering (cf., CHERI). It also reduces the amount of metadata that must be stored at execution time by requiring only storage for rules pertaining to the active set of execution contexts. Thanks to limiting the rules that are needed for enforcement at any given time, \SHORTNAME also makes it feasible to cache that metadata in on-chip memory, and enable implicit access checks with no overhead.

\section{Conclusion}
\label{sec:conclusion}

By implementing and evaluating \SHORTNAME, we demonstrated that \ABBRNAME is a novel, effective approach to protect against memory vulnerabilities at run time. \SHORTNAME can also enforce memory isolation at coarser or finer granularity, to facilitate different types of memory protection strategies. 
In future work we plan to integrate \SHORTNAME with a general purpose RISC-V core (e.g. the Rocket Core~\cite{Asanovic17} and extend our \ABBRNAME GCC Plug-in to support more protection models.
To support reproducibility of our results, we provide 1) our enhanced GCC compiler, which can automatically instrument unmodified C programs; 2) instrumented binaries of our test programs; and 3) a version of the official RISC-V simulator with added support for \SHORTNAME instructions.

\ifnotanonymous
\section*{Acknowledgements}
This work was supported by the German Science Foundation CRC 1119 CROSSING project S2, the German Federal Ministry of Education and Research (BMBF) within CRISP, the EU's Horizon 2020 research and innovation program under grant nr. 643964 (SUPERCLOUD), Tekes --- the Finnish Funding Agency for Innovation under grant nr. 3881/31/2016 (CloSer), Academy of Finland under grant nr. 309994 (SELIoT), and the Intel Collaborative Research Institute for Secure Computing (ICRI-SC).

The authors would also like to extend their thanks the Aalto University Secure Systems Group interns Kesara Gamlath  and Rangana De Silva who provided assistance in the efforts to realize and test \SHORTNAME on the FPGA.

\fi

\ifabridged
{\footnotesize \bibliographystyle{acm}
\else
\bibliographystyle{ACM-Reference-Format}
\fi
\bibliography{hw-enforced-scoping} 


\begin{thebibliography}{00}

%%% the Web address.

\ifx \showCODEN    \undefined \def \showCODEN     #1{\unskip}     \fi
\ifx \showDOI      \undefined \def \showDOI       #1{{\tt DOI:}\penalty0{#1}\ }
  \fi
\ifx \showISBNx    \undefined \def \showISBNx     #1{\unskip}     \fi
\ifx \showISBNxiii \undefined \def \showISBNxiii  #1{\unskip}     \fi
\ifx \showISSN     \undefined \def \showISSN      #1{\unskip}     \fi
\ifx \showLCCN     \undefined \def \showLCCN      #1{\unskip}     \fi
\ifx \shownote     \undefined \def \shownote      #1{#1}          \fi
\ifx \showarticletitle \undefined \def \showarticletitle #1{#1}   \fi
\ifx \showURL      \undefined \def \showURL       {\relax}        \fi
% invisible to TeX
\providecommand\bibfield[2]{#2}
\providecommand\bibinfo[2]{#2}
\providecommand\natexlab[1]{#1}
\providecommand\showeprint[2][]{arXiv:#2}

\bibitem[\protect\citeauthoryear{Abadi, Budiu, Erlingsson, and Ligatti}{Abadi
  et~al\mbox{.}}{2009}]%
        {Abadi09}
\bibfield{author}{\bibinfo{person}{Mart\'{\i}n Abadi}, \bibinfo{person}{Mihai
  Budiu}, \bibinfo{person}{\'{U}lfar Erlingsson}, {and} \bibinfo{person}{Jay
  Ligatti}.} \bibinfo{year}{2009}\natexlab{}.
\newblock \showarticletitle{Control-flow Integrity Principles, Implementations,
  and Applications}.
\newblock \bibinfo{journal}{{\em ACM Trans. Inf. Syst. Secur.\/}}
  \bibinfo{volume}{13}, \bibinfo{number}{1}, Article \bibinfo{articleno}{4}
  (\bibinfo{date}{Nov.} \bibinfo{year}{2009}), \bibinfo{numpages}{40}~pages.
\newblock
\showISSN{1094-9224}
\showDOI{%
\url{https://doi.org/10.1145/1609956.1609960}}


\bibitem[\protect\citeauthoryear{Abera, Asokan, Davi, Ekberg, Nyman, Paverd,
  Sadeghi, and Tsudik}{Abera et~al\mbox{.}}{2016}]%
        {Abera16}
\bibfield{author}{\bibinfo{person}{Tigist Abera}, \bibinfo{person}{N. Asokan},
  \bibinfo{person}{Lucas Davi}, \bibinfo{person}{Jan-Erik Ekberg},
  \bibinfo{person}{Thomas Nyman}, \bibinfo{person}{Andrew Paverd},
  \bibinfo{person}{Ahmad-Reza Sadeghi}, {and} \bibinfo{person}{Gene Tsudik}.}
  \bibinfo{year}{2016}\natexlab{}.
\newblock \showarticletitle{C-FLAT: Control-Flow Attestation for Embedded
  Systems Software}. In \bibinfo{booktitle}{{\em Proceedings of the 2016 ACM
  SIGSAC Conference on Computer and Communications Security}} {\em
  (\bibinfo{series}{CCS '16})}. \bibinfo{publisher}{ACM}, \bibinfo{address}{New
  York, NY, USA}, \bibinfo{pages}{743--754}.
\newblock
\showISBNx{978-1-4503-4139-4}
\showDOI{%
\url{https://doi.org/10.1145/2976749.2978358}}


\bibitem[\protect\citeauthoryear{Akritidis, Cadar, Raiciu, Costa, and
  Castro}{Akritidis et~al\mbox{.}}{2008}]%
        {Akritidis08}
\bibfield{author}{\bibinfo{person}{Periklis Akritidis},
  \bibinfo{person}{Cristian Cadar}, \bibinfo{person}{Costin Raiciu},
  \bibinfo{person}{Manuel Costa}, {and} \bibinfo{person}{Miguel Castro}.}
  \bibinfo{year}{2008}\natexlab{}.
\newblock \showarticletitle{Preventing Memory Error Exploits with WIT}. In
  \bibinfo{booktitle}{{\em Proceedings of the 2008 IEEE Symposium on Security
  and Privacy}} {\em (\bibinfo{series}{SP '08})}. \bibinfo{publisher}{IEEE
  Computer Society}, \bibinfo{address}{Washington, DC, USA},
  \bibinfo{pages}{263--277}.
\newblock
\showISBNx{978-0-7695-3168-7}
\showDOI{%
\url{https://doi.org/10.1109/SP.2008.30}}


\bibitem[\protect\citeauthoryear{Asanović, Avizienis, Bachrach, Beamer,
  Biancolin, Celio, Cook, Dabbelt, Hauser, Izraelevitz, Karandikar, Keller,
  Kim, Koenig, Lee, Love, Maas, Magyar, Mao, Moreto, Ou, Patterson, Richards,
  Schmidt, Twigg, Vo, and Waterman}{Asanović et~al\mbox{.}}{2016}]%
        {Asanovic17}
\bibfield{author}{\bibinfo{person}{Krste Asanović}, \bibinfo{person}{Rimas
  Avizienis}, \bibinfo{person}{Jonathan Bachrach}, \bibinfo{person}{Scott
  Beamer}, \bibinfo{person}{David Biancolin}, \bibinfo{person}{Christopher
  Celio}, \bibinfo{person}{Henry Cook}, \bibinfo{person}{Daniel Dabbelt},
  \bibinfo{person}{John Hauser}, \bibinfo{person}{Adam Izraelevitz},
  \bibinfo{person}{Sagar Karandikar}, \bibinfo{person}{Ben Keller},
  \bibinfo{person}{Donggyu Kim}, \bibinfo{person}{John Koenig},
  \bibinfo{person}{Yunsup Lee}, \bibinfo{person}{Eric Love},
  \bibinfo{person}{Martin Maas}, \bibinfo{person}{Albert Magyar},
  \bibinfo{person}{Howard Mao}, \bibinfo{person}{Miquel Moreto},
  \bibinfo{person}{Albert Ou}, \bibinfo{person}{David~A. Patterson},
  \bibinfo{person}{Brian Richards}, \bibinfo{person}{Colin Schmidt},
  \bibinfo{person}{Stephen Twigg}, \bibinfo{person}{Huy Vo}, {and}
  \bibinfo{person}{Andrew Waterman}.} \bibinfo{year}{2016}\natexlab{}.
\newblock \bibinfo{booktitle}{{\em The Rocket Chip Generator}}.
\newblock \bibinfo{type}{{T}echnical {R}eport} UCB/EECS-2016-17.
  \bibinfo{institution}{EECS Department, University of California, Berkeley}.
\newblock
\showURL{%
\url{http://www2.eecs.berkeley.edu/Pubs/TechRpts/2016/EECS-2016-17.html}}


\bibitem[\protect\citeauthoryear{Bhatkar and Sekar}{Bhatkar and Sekar}{2008}]%
        {Bhatkar08}
\bibfield{author}{\bibinfo{person}{Sandeep Bhatkar} {and} \bibinfo{person}{R.
  Sekar}.} \bibinfo{year}{2008}\natexlab{}.
\newblock \showarticletitle{Data Space Randomization}. In
  \bibinfo{booktitle}{{\em Proceedings of the 5th International Conference on
  Detection of Intrusions and Malware, and Vulnerability Assessment}} {\em
  (\bibinfo{series}{DIMVA '08})}. \bibinfo{publisher}{Springer-Verlag},
  \bibinfo{address}{Berlin, Heidelberg}, \bibinfo{pages}{1--22}.
\newblock
\showISBNx{978-3-540-70541-3}
\showDOI{%
\url{https://doi.org/10.1007/978-3-540-70542-0_1}}


\bibitem[\protect\citeauthoryear{Cadar, Akritidis, Costa, Martin, and
  Castro}{Cadar et~al\mbox{.}}{2008}]%
        {Cadar08}
\bibfield{author}{\bibinfo{person}{Cristian Cadar}, \bibinfo{person}{Periklis
  Akritidis}, \bibinfo{person}{Manuel Costa}, \bibinfo{person}{Jean-Philippe
  Martin}, {and} \bibinfo{person}{Miguel Castro}.}
  \bibinfo{year}{2008}\natexlab{}.
\newblock \bibinfo{booktitle}{{\em Data Randomization}}.
\newblock \bibinfo{type}{{T}echnical {R}eport}. \bibinfo{pages}{14} pages.
\newblock
\showURL{%
\url{https://www.microsoft.com/en-us/research/publication/data-randomization/}}


\bibitem[\protect\citeauthoryear{Castro, Costa, and Harris}{Castro
  et~al\mbox{.}}{2006}]%
        {Castro06}
\bibfield{author}{\bibinfo{person}{Miguel Castro}, \bibinfo{person}{Manuel
  Costa}, {and} \bibinfo{person}{Tim Harris}.} \bibinfo{year}{2006}\natexlab{}.
\newblock \showarticletitle{Securing Software by Enforcing Data-flow
  Integrity}. In \bibinfo{booktitle}{{\em Proceedings of the 7th Symposium on
  Operating Systems Design and Implementation}} {\em (\bibinfo{series}{OSDI
  '06})}. \bibinfo{publisher}{USENIX Association}, \bibinfo{address}{Berkeley,
  CA, USA}, \bibinfo{pages}{147--160}.
\newblock
\showISBNx{1-931971-47-1}
\showURL{%
\url{http://dl.acm.org/citation.cfm?id=1298455.1298470}}


\bibitem[\protect\citeauthoryear{Castro, Costa, Martin, Peinado, Akritidis,
  Donnelly, Barham, and Black}{Castro et~al\mbox{.}}{2009}]%
        {Castro09}
\bibfield{author}{\bibinfo{person}{Miguel Castro}, \bibinfo{person}{Manuel
  Costa}, \bibinfo{person}{Jean-Philippe Martin}, \bibinfo{person}{Marcus
  Peinado}, \bibinfo{person}{Periklis Akritidis}, \bibinfo{person}{Austin
  Donnelly}, \bibinfo{person}{Paul Barham}, {and} \bibinfo{person}{Richard
  Black}.} \bibinfo{year}{2009}\natexlab{}.
\newblock \showarticletitle{Fast Byte-granularity Software Fault Isolation}. In
  \bibinfo{booktitle}{{\em Proceedings of the ACM SIGOPS 22Nd Symposium on
  Operating Systems Principles}} {\em (\bibinfo{series}{SOSP '09})}.
  \bibinfo{publisher}{ACM}, \bibinfo{address}{New York, NY, USA},
  \bibinfo{pages}{45--58}.
\newblock
\showISBNx{978-1-60558-752-3}
\showDOI{%
\url{https://doi.org/10.1145/1629575.1629581}}


\bibitem[\protect\citeauthoryear{Cowan, Beattie, Johansen, and Wagle}{Cowan
  et~al\mbox{.}}{2003}]%
        {Cowan03}
\bibfield{author}{\bibinfo{person}{Crispin Cowan}, \bibinfo{person}{Steve
  Beattie}, \bibinfo{person}{John Johansen}, {and} \bibinfo{person}{Perry
  Wagle}.} \bibinfo{year}{2003}\natexlab{}.
\newblock \showarticletitle{Pointguard: Protecting Pointers from Buffer
  Overflow Vulnerabilities}. In \bibinfo{booktitle}{{\em Proceedings of the
  12th Conference on USENIX Security Symposium - Volume 12}} {\em
  (\bibinfo{series}{SSYM'03})}. \bibinfo{publisher}{USENIX Association},
  \bibinfo{address}{Berkeley, CA, USA}, \bibinfo{pages}{7--7}.
\newblock
\showURL{%
\url{http://dl.acm.org/citation.cfm?id=1251353.1251360}}


\bibitem[\protect\citeauthoryear{Dessouky, Zeitouni, Nyman, Paverd, Davi,
  Koeberl, Asokan, and Sadeghi}{Dessouky et~al\mbox{.}}{2017}]%
        {Dessouky17}
\bibfield{author}{\bibinfo{person}{Ghada Dessouky}, \bibinfo{person}{Shaza
  Zeitouni}, \bibinfo{person}{Thomas Nyman}, \bibinfo{person}{Andrew Paverd},
  \bibinfo{person}{Lucas Davi}, \bibinfo{person}{Patrick Koeberl},
  \bibinfo{person}{N. Asokan}, {and} \bibinfo{person}{Ahmad-Reza Sadeghi}.}
  \bibinfo{year}{2017}\natexlab{}.
\newblock \showarticletitle{LO-FAT: Low-Overhead Control Flow ATtestation in
  Hardware}. In \bibinfo{booktitle}{{\em Proceedings of the 54th Annual Design
  Automation Conference 2017}} {\em (\bibinfo{series}{DAC '17})}.
  \bibinfo{publisher}{ACM}, \bibinfo{address}{New York, NY, USA}, Article
  \bibinfo{articleno}{24}, \bibinfo{numpages}{6}~pages.
\newblock
\showISBNx{978-1-4503-4927-7}
\showDOI{%
\url{https://doi.org/10.1145/3061639.3062276}}


\bibitem[\protect\citeauthoryear{Devietti, Blundell, Martin, and
  Zdancewic}{Devietti et~al\mbox{.}}{2008}]%
        {Devietti08}
\bibfield{author}{\bibinfo{person}{Joe Devietti}, \bibinfo{person}{Colin
  Blundell}, \bibinfo{person}{Milo M.~K. Martin}, {and} \bibinfo{person}{Steve
  Zdancewic}.} \bibinfo{year}{2008}\natexlab{}.
\newblock \showarticletitle{Hardbound: Architectural Support for Spatial Safety
  of the C Programming Language}. In \bibinfo{booktitle}{{\em Proceedings of
  the 13th International Conference on Architectural Support for Programming
  Languages and Operating Systems}} {\em (\bibinfo{series}{ASPLOS XIII})}.
  \bibinfo{publisher}{ACM}, \bibinfo{address}{New York, NY, USA},
  \bibinfo{pages}{103--114}.
\newblock
\showISBNx{978-1-59593-958-6}
\showDOI{%
\url{https://doi.org/10.1145/1346281.1346295}}


\bibitem[\protect\citeauthoryear{Duck, Yap, and Cavallaro}{Duck
  et~al\mbox{.}}{2017}]%
        {Duck17}
\bibfield{author}{\bibinfo{person}{Gregory~J Duck}, \bibinfo{person}{Roland~HC
  Yap}, {and} \bibinfo{person}{Lorenzo Cavallaro}.}
  \bibinfo{year}{2017}\natexlab{}.
\newblock \showarticletitle{Stack Bounds Protection with Low Fat Pointers}.
\newblock  (\bibinfo{year}{2017}).
\newblock


\bibitem[\protect\citeauthoryear{Erlingsson, Abadi, Vrable, Budiu, and
  Necula}{Erlingsson et~al\mbox{.}}{2006}]%
        {Erlingsson06}
\bibfield{author}{\bibinfo{person}{\'{U}lfar Erlingsson},
  \bibinfo{person}{Mart\'{\i}n Abadi}, \bibinfo{person}{Michael Vrable},
  \bibinfo{person}{Mihai Budiu}, {and} \bibinfo{person}{George~C. Necula}.}
  \bibinfo{year}{2006}\natexlab{}.
\newblock \showarticletitle{XFI: Software Guards for System Address Spaces}. In
  \bibinfo{booktitle}{{\em Proceedings of the 7th Symposium on Operating
  Systems Design and Implementation}} {\em (\bibinfo{series}{OSDI '06})}.
  \bibinfo{publisher}{USENIX Association}, \bibinfo{address}{Berkeley, CA,
  USA}, \bibinfo{pages}{75--88}.
\newblock
\showISBNx{1-931971-47-1}
\showURL{%
\url{http://dl.acm.org/citation.cfm?id=1298455.1298463}}


\bibitem[\protect\citeauthoryear{Evans}{Evans}{2016}]%
        {Evans16}
\bibfield{author}{\bibinfo{person}{Chris Evans}.}
  \bibinfo{year}{2016}\natexlab{}.
\newblock \bibinfo{title}{Advancing exploitation: a scriptless 0day exploit
  against Linux desktops}.
\newblock   (\bibinfo{year}{2016}).
\newblock
\showURL{%
\url{https://scarybeastsecurity.blogspot.com/2016/11/0day-exploit-advancing-exploitation.html}}


\bibitem[\protect\citeauthoryear{Group, IEEE, EU, and Council}{Group
  et~al\mbox{.}}{2017}]%
        {IoTDeveloperSurvey17}
\bibfield{author}{\bibinfo{person}{Eclipse IoT~Working Group},
  \bibinfo{person}{IEEE}, \bibinfo{person}{Agile-IoT EU}, {and}
  \bibinfo{person}{IoT Council}.} \bibinfo{year}{2017}\natexlab{}.
\newblock \bibinfo{title}{{IoT Developer Survey 2017}}.
\newblock   (\bibinfo{year}{2017}).
\newblock
\showURL{%
\url{https://ianskerrett.wordpress.com/2017/04/19/iot-developer-trends-2017-edition/}}


\bibitem[\protect\citeauthoryear{Hewlett-Packard}{Hewlett-Packard}{2006}]%
        {HP-DEP}
\bibfield{author}{\bibinfo{person}{Hewlett-Packard}.}
  \bibinfo{year}{2006}\natexlab{}.
\newblock \bibinfo{title}{{Data Execution Prevention}}.
\newblock
  \bibinfo{howpublished}{\url{http://h10032.www1.hp.com/ctg/Manual/c00387685.pdf}}.
    (\bibinfo{year}{2006}).
\newblock


\bibitem[\protect\citeauthoryear{Hu, Shinde, Adrian, Chua, Saxena, and
  Liang}{Hu et~al\mbox{.}}{2016}]%
        {Hu16}
\bibfield{author}{\bibinfo{person}{Hong Hu}, \bibinfo{person}{Shweta Shinde},
  \bibinfo{person}{Sendroiu Adrian}, \bibinfo{person}{Zheng~Leong Chua},
  \bibinfo{person}{Prateek Saxena}, {and} \bibinfo{person}{Zhenkai Liang}.}
  \bibinfo{year}{2016}\natexlab{}.
\newblock \showarticletitle{Data-Oriented Programming: On the Expressiveness of
  Non-control Data Attacks}. In \bibinfo{booktitle}{{\em {IEEE} Symposium on
  Security and Privacy, {SP} 2016, San Jose, CA, USA, May 22-26, 2016}}.
  \bibinfo{pages}{969--986}.
\newblock
\showDOI{%
\url{https://doi.org/10.1109/SP.2016.62}}


\bibitem[\protect\citeauthoryear{Ibrahim and Rupp}{Ibrahim and Rupp}{2013}]%
        {Ibrahim13}
\bibfield{author}{\bibinfo{person}{Mostafa~E.A. Ibrahim} {and}
  \bibinfo{person}{Markus Rupp}.} \bibinfo{year}{2013}\natexlab{}.
\newblock \showarticletitle{Embedded Systems Code Optimization and Power
  Consumption}.
\newblock In \bibinfo{booktitle}{{\em Embedded and Networking Systems: Design,
  Software, and Implementation}}, \bibfield{editor}{\bibinfo{person}{Gul~N.
  Khan} {and} \bibinfo{person}{Krzysztof Iniewski}} (Eds.).
  \bibinfo{publisher}{Chapman and Hall/CRC}, \bibinfo{address}{Boca Raton,
  Florida, United States}, Chapter~4, \bibinfo{pages}{85--103}.
\newblock


\bibitem[\protect\citeauthoryear{Intel}{Intel}{2016}]%
        {Intel-CET}
\bibfield{author}{\bibinfo{person}{Intel}.} \bibinfo{year}{2016}\natexlab{}.
\newblock \bibinfo{title}{{Control-flow Enforcement Technology Preview}}.
\newblock   (\bibinfo{year}{2016}).
\newblock
\showURL{%
\url{https://software.intel.com/sites/default/files/managed/4d/2a/control-flow-enforcement-technology-preview.pdf}}


\bibitem[\protect\citeauthoryear{Kuvaiskii, Oleksenko, Arnautov, Trach,
  Bhatotia, Felber, and Fetzer}{Kuvaiskii et~al\mbox{.}}{2017}]%
        {Kuvaiskii17}
\bibfield{author}{\bibinfo{person}{Dmitrii Kuvaiskii}, \bibinfo{person}{Oleksii
  Oleksenko}, \bibinfo{person}{Sergei Arnautov}, \bibinfo{person}{Bohdan
  Trach}, \bibinfo{person}{Pramod Bhatotia}, \bibinfo{person}{Pascal Felber},
  {and} \bibinfo{person}{Christof Fetzer}.} \bibinfo{year}{2017}\natexlab{}.
\newblock \showarticletitle{SGXBOUNDS: Memory Safety for Shielded Execution}.
  In \bibinfo{booktitle}{{\em Proceedings of the Twelfth European Conference on
  Computer Systems}} {\em (\bibinfo{series}{EuroSys '17})}.
  \bibinfo{publisher}{ACM}, \bibinfo{address}{New York, NY, USA},
  \bibinfo{pages}{205--221}.
\newblock
\showISBNx{978-1-4503-4938-3}
\showDOI{%
\url{https://doi.org/10.1145/3064176.3064192}}


\bibitem[\protect\citeauthoryear{Kuznetsov, Szekeres, Payer, Candea, Sekar, and
  Song}{Kuznetsov et~al\mbox{.}}{2014}]%
        {Kuznetsov14}
\bibfield{author}{\bibinfo{person}{Volodymyr Kuznetsov},
  \bibinfo{person}{L\'{a}szl\'{o} Szekeres}, \bibinfo{person}{Mathias Payer},
  \bibinfo{person}{George Candea}, \bibinfo{person}{R. Sekar}, {and}
  \bibinfo{person}{Dawn Song}.} \bibinfo{year}{2014}\natexlab{}.
\newblock \showarticletitle{Code-pointer Integrity}. In
  \bibinfo{booktitle}{{\em Proceedings of the 11th USENIX Conference on
  Operating Systems Design and Implementation}} {\em
  (\bibinfo{series}{OSDI'14})}. \bibinfo{publisher}{USENIX Association},
  \bibinfo{address}{Berkeley, CA, USA}, \bibinfo{pages}{147--163}.
\newblock
\showISBNx{978-1-931971-16-4}
\showURL{%
\url{http://dl.acm.org/citation.cfm?id=2685048.2685061}}


\bibitem[\protect\citeauthoryear{Kwon, Dhawan, Smith, Knight, and DeHon}{Kwon
  et~al\mbox{.}}{2013}]%
        {Kwon13}
\bibfield{author}{\bibinfo{person}{Albert Kwon}, \bibinfo{person}{Udit Dhawan},
  \bibinfo{person}{Jonathan~M. Smith}, \bibinfo{person}{Thomas~F. Knight, Jr.},
  {and} \bibinfo{person}{Andre DeHon}.} \bibinfo{year}{2013}\natexlab{}.
\newblock \showarticletitle{Low-fat Pointers: Compact Encoding and Efficient
  Gate-level Implementation of Fat Pointers for Spatial Safety and
  Capability-based Security}. In \bibinfo{booktitle}{{\em Proceedings of the
  2013 ACM SIGSAC Conference on Computer \&\#38; Communications Security}} {\em
  (\bibinfo{series}{CCS '13})}. \bibinfo{publisher}{ACM}, \bibinfo{address}{New
  York, NY, USA}, \bibinfo{pages}{721--732}.
\newblock
\showISBNx{978-1-4503-2477-9}
\showDOI{%
\url{https://doi.org/10.1145/2508859.2516713}}


\bibitem[\protect\citeauthoryear{Larsen, Homescu, Brunthaler, and Franz}{Larsen
  et~al\mbox{.}}{2014}]%
        {Larsen14}
\bibfield{author}{\bibinfo{person}{Per Larsen}, \bibinfo{person}{Andrei
  Homescu}, \bibinfo{person}{Stefan Brunthaler}, {and} \bibinfo{person}{Michael
  Franz}.} \bibinfo{year}{2014}\natexlab{}.
\newblock \showarticletitle{SoK: Automated Software Diversity}. In
  \bibinfo{booktitle}{{\em Proceedings of the 2014 IEEE Symposium on Security
  and Privacy}} {\em (\bibinfo{series}{SP '14})}. \bibinfo{publisher}{IEEE
  Computer Society}, \bibinfo{address}{Washington, DC, USA},
  \bibinfo{pages}{276--291}.
\newblock
\showISBNx{978-1-4799-4686-0}
\showDOI{%
\url{https://doi.org/10.1109/SP.2014.25}}


\bibitem[\protect\citeauthoryear{Mao, Chen, Zhou, Wang, Zeldovich, and
  Kaashoek}{Mao et~al\mbox{.}}{2011}]%
        {Mao11}
\bibfield{author}{\bibinfo{person}{Yandong Mao}, \bibinfo{person}{Haogang
  Chen}, \bibinfo{person}{Dong Zhou}, \bibinfo{person}{Xi Wang},
  \bibinfo{person}{Nickolai Zeldovich}, {and} \bibinfo{person}{M.~Frans
  Kaashoek}.} \bibinfo{year}{2011}\natexlab{}.
\newblock \showarticletitle{Software Fault Isolation with API Integrity and
  Multi-principal Modules}. In \bibinfo{booktitle}{{\em Proceedings of the
  Twenty-Third ACM Symposium on Operating Systems Principles}} {\em
  (\bibinfo{series}{SOSP '11})}. \bibinfo{publisher}{ACM},
  \bibinfo{address}{New York, NY, USA}, \bibinfo{pages}{115--128}.
\newblock
\showISBNx{978-1-4503-0977-6}
\showDOI{%
\url{https://doi.org/10.1145/2043556.2043568}}


\bibitem[\protect\citeauthoryear{Nagarakatte, Zhao, Martin, and
  Zdancewic}{Nagarakatte et~al\mbox{.}}{2009}]%
        {Nagarakatte09}
\bibfield{author}{\bibinfo{person}{Santosh Nagarakatte},
  \bibinfo{person}{Jianzhou Zhao}, \bibinfo{person}{M.K. Martin, Milo}, {and}
  \bibinfo{person}{Steve Zdancewic}.} \bibinfo{year}{2009}\natexlab{}.
\newblock \showarticletitle{SoftBound: Highly Compatible and Complete Spatial
  Memory Safety for C}. In \bibinfo{booktitle}{{\em Proceedings of the 30th ACM
  SIGPLAN Conference on Programming Language Design and Implementation}} {\em
  (\bibinfo{series}{PLDI '09})}. \bibinfo{publisher}{ACM},
  \bibinfo{address}{New York, NY, USA}, \bibinfo{pages}{245--258}.
\newblock
\showISBNx{978-1-60558-392-1}
\showDOI{%
\url{https://doi.org/10.1145/1542476.1542504}}


\bibitem[\protect\citeauthoryear{Nagarakatte, Zhao, Martin, and
  Zdancewic}{Nagarakatte et~al\mbox{.}}{2010}]%
        {Nagarakatte10}
\bibfield{author}{\bibinfo{person}{Santosh Nagarakatte},
  \bibinfo{person}{Jianzhou Zhao}, \bibinfo{person}{M.K. Martin, Milo}, {and}
  \bibinfo{person}{Steve Zdancewic}.} \bibinfo{year}{2010}\natexlab{}.
\newblock \showarticletitle{CETS: Compiler Enforced Temporal Safety for C}. In
  \bibinfo{booktitle}{{\em Proceedings of the 2010 International Symposium on
  Memory Management}} {\em (\bibinfo{series}{ISMM '10})}.
  \bibinfo{publisher}{ACM}, \bibinfo{address}{New York, NY, USA},
  \bibinfo{pages}{31--40}.
\newblock
\showISBNx{978-1-4503-0054-4}
\showDOI{%
\url{https://doi.org/10.1145/1806651.1806657}}


\bibitem[\protect\citeauthoryear{Necula, McPeak, and Weimer}{Necula
  et~al\mbox{.}}{2002}]%
        {Necula02}
\bibfield{author}{\bibinfo{person}{George~C. Necula}, \bibinfo{person}{Scott
  McPeak}, {and} \bibinfo{person}{Westley Weimer}.}
  \bibinfo{year}{2002}\natexlab{}.
\newblock \showarticletitle{CCured: Type-safe Retrofitting of Legacy Code}. In
  \bibinfo{booktitle}{{\em Proceedings of the 29th ACM SIGPLAN-SIGACT Symposium
  on Principles of Programming Languages}} {\em (\bibinfo{series}{POPL '02})}.
  \bibinfo{publisher}{ACM}, \bibinfo{address}{New York, NY, USA},
  \bibinfo{pages}{128--139}.
\newblock
\showISBNx{1-58113-450-9}
\showDOI{%
\url{https://doi.org/10.1145/503272.503286}}


\bibitem[\protect\citeauthoryear{Oleksenko, Kuvaiskii, Bhatotia, Felber, and
  Fetzer}{Oleksenko et~al\mbox{.}}{2017}]%
        {Oleksenko17}
\bibfield{author}{\bibinfo{person}{Oleksii Oleksenko}, \bibinfo{person}{Dmitrii
  Kuvaiskii}, \bibinfo{person}{Pramod Bhatotia}, \bibinfo{person}{Pascal
  Felber}, {and} \bibinfo{person}{Christof Fetzer}.}
  \bibinfo{year}{2017}\natexlab{}.
\newblock \showarticletitle{Intel {MPX} Explained: An Empirical Study of Intel
  {MPX} and Software-based Bounds Checking Approaches}.
\newblock \bibinfo{journal}{{\em CoRR\/}}  \bibinfo{volume}{abs/1702.00719}
  (\bibinfo{year}{2017}).
\newblock
\showURL{%
\url{http://arxiv.org/abs/1702.00719}}


\bibitem[\protect\citeauthoryear{{Qualcomm Technologies Inc.}}{{Qualcomm
  Technologies Inc.}}{2017}]%
        {PAC}
\bibfield{author}{\bibinfo{person}{{Qualcomm Technologies Inc.}}}
  \bibinfo{year}{2017}\natexlab{}.
\newblock \bibinfo{title}{{Pointer Authentication on ARMv8.3}}.
\newblock
  \bibinfo{howpublished}{\url{https://www.qualcomm.com/media/documents/files/whitepaper-pointer-authentication-on-armv8-3.pdf}}.
    (\bibinfo{year}{2017}).
\newblock


\bibitem[\protect\citeauthoryear{Schlesinger, Pattabiraman, Swamy, Walker, and
  Zorn}{Schlesinger et~al\mbox{.}}{2011}]%
        {Schlesinger11}
\bibfield{author}{\bibinfo{person}{C. Schlesinger}, \bibinfo{person}{K.
  Pattabiraman}, \bibinfo{person}{N. Swamy}, \bibinfo{person}{D. Walker}, {and}
  \bibinfo{person}{B. Zorn}.} \bibinfo{year}{2011}\natexlab{}.
\newblock \showarticletitle{Modular Protections against Non-control Data
  Attacks}. In \bibinfo{booktitle}{{\em 2011 IEEE 24th Computer Security
  Foundations Symposium}}. \bibinfo{pages}{131--145}.
\newblock
\showISSN{1063-6900}
\showDOI{%
\url{https://doi.org/10.1109/CSF.2011.16}}


\bibitem[\protect\citeauthoryear{Serebryany, Bruening, Potapenko, and
  Vyukov}{Serebryany et~al\mbox{.}}{2012}]%
        {Serebryany12}
\bibfield{author}{\bibinfo{person}{Konstantin Serebryany},
  \bibinfo{person}{Derek Bruening}, \bibinfo{person}{Alexander Potapenko},
  {and} \bibinfo{person}{Dmitriy Vyukov}.} \bibinfo{year}{2012}\natexlab{}.
\newblock \showarticletitle{AddressSanitizer: A Fast Address Sanity Checker}.
  In \bibinfo{booktitle}{{\em Presented as part of the 2012 USENIX Annual
  Technical Conference (USENIX ATC 12)}}. \bibinfo{publisher}{USENIX},
  \bibinfo{address}{Boston, MA}, \bibinfo{pages}{309--318}.
\newblock
\showISBNx{978-931971-93-5}
\showURL{%
\url{https://www.usenix.org/conference/atc12/technical-sessions/presentation/serebryany}}


\bibitem[\protect\citeauthoryear{Shacham}{Shacham}{2007}]%
        {Shacham07}
\bibfield{author}{\bibinfo{person}{Hovav Shacham}.}
  \bibinfo{year}{2007}\natexlab{}.
\newblock \showarticletitle{The Geometry of Innocent Flesh on the Bone:
  Return-into-libc Without Function Calls (on the x86)}. In
  \bibinfo{booktitle}{{\em Proceedings of the 14th ACM Conference on Computer
  and Communications Security}} {\em (\bibinfo{series}{CCS '07})}.
  \bibinfo{publisher}{ACM}, \bibinfo{address}{New York, NY, USA},
  \bibinfo{pages}{552--561}.
\newblock
\showISBNx{978-1-59593-703-2}
\showDOI{%
\url{https://doi.org/10.1145/1315245.1315313}}


\bibitem[\protect\citeauthoryear{Shacham, Page, Pfaff, Goh, Modadugu, and
  Boneh}{Shacham et~al\mbox{.}}{2004}]%
        {Shacham04}
\bibfield{author}{\bibinfo{person}{Hovav Shacham}, \bibinfo{person}{Matthew
  Page}, \bibinfo{person}{Ben Pfaff}, \bibinfo{person}{Eu-Jin Goh},
  \bibinfo{person}{Nagendra Modadugu}, {and} \bibinfo{person}{Dan Boneh}.}
  \bibinfo{year}{2004}\natexlab{}.
\newblock \showarticletitle{On the Effectiveness of Address-space
  Randomization}. In \bibinfo{booktitle}{{\em Proceedings of the 11th ACM
  Conference on Computer and Communications Security}} {\em
  (\bibinfo{series}{CCS '04})}. \bibinfo{publisher}{ACM}, \bibinfo{address}{New
  York, NY, USA}, \bibinfo{pages}{298--307}.
\newblock
\showISBNx{1-58113-961-6}
\showDOI{%
\url{https://doi.org/10.1145/1030083.1030124}}


\bibitem[\protect\citeauthoryear{Song, Moon, Alam, Yun, Lee, Kim, Lee, and
  Paek}{Song et~al\mbox{.}}{2016}]%
        {Song16}
\bibfield{author}{\bibinfo{person}{C. Song}, \bibinfo{person}{H. Moon},
  \bibinfo{person}{M. Alam}, \bibinfo{person}{I. Yun}, \bibinfo{person}{B.
  Lee}, \bibinfo{person}{T. Kim}, \bibinfo{person}{W. Lee}, {and}
  \bibinfo{person}{Y. Paek}.} \bibinfo{year}{2016}\natexlab{}.
\newblock \showarticletitle{HDFI: Hardware-Assisted Data-Flow Isolation}. In
  \bibinfo{booktitle}{{\em 2016 IEEE Symposium on Security and Privacy (SP)}}.
  \bibinfo{pages}{1--17}.
\newblock
\showDOI{%
\url{https://doi.org/10.1109/SP.2016.9}}


\bibitem[\protect\citeauthoryear{Wahbe, Lucco, Anderson, and Graham}{Wahbe
  et~al\mbox{.}}{1993}]%
        {Wahbe93a}
\bibfield{author}{\bibinfo{person}{Robert Wahbe}, \bibinfo{person}{Steven
  Lucco}, \bibinfo{person}{Thomas~E. Anderson}, {and} \bibinfo{person}{Susan~L.
  Graham}.} \bibinfo{year}{1993}\natexlab{}.
\newblock \showarticletitle{Efficient Software-based Fault Isolation}. In
  \bibinfo{booktitle}{{\em Proceedings of the Fourteenth ACM Symposium on
  Operating Systems Principles}} {\em (\bibinfo{series}{SOSP '93})}.
  \bibinfo{publisher}{ACM}, \bibinfo{address}{New York, NY, USA},
  \bibinfo{pages}{203--216}.
\newblock
\showISBNx{0-89791-632-8}
\showDOI{%
\url{https://doi.org/10.1145/168619.168635}}


\bibitem[\protect\citeauthoryear{Watson, Woodruff, Neumann, Moore, Anderson,
  Chisnall, Dave, Davis, Gudka, Laurie, Murdoch, Norton, Roe, Son, and
  Vadera}{Watson et~al\mbox{.}}{2015}]%
        {Watson15}
\bibfield{author}{\bibinfo{person}{R.~N.~M. Watson}, \bibinfo{person}{J.
  Woodruff}, \bibinfo{person}{P.~G. Neumann}, \bibinfo{person}{S.~W. Moore},
  \bibinfo{person}{J. Anderson}, \bibinfo{person}{D. Chisnall},
  \bibinfo{person}{N. Dave}, \bibinfo{person}{B. Davis}, \bibinfo{person}{K.
  Gudka}, \bibinfo{person}{B. Laurie}, \bibinfo{person}{S.~J. Murdoch},
  \bibinfo{person}{R. Norton}, \bibinfo{person}{M. Roe}, \bibinfo{person}{S.
  Son}, {and} \bibinfo{person}{M. Vadera}.} \bibinfo{year}{2015}\natexlab{}.
\newblock \showarticletitle{CHERI: A Hybrid Capability-System Architecture for
  Scalable Software Compartmentalization}. In \bibinfo{booktitle}{{\em 2015
  IEEE Symposium on Security and Privacy}}. \bibinfo{pages}{20--37}.
\newblock
\showISSN{1081-6011}
\showDOI{%
\url{https://doi.org/10.1109/SP.2015.9}}


\bibitem[\protect\citeauthoryear{Woodruff, Watson, Chisnall, Moore, Anderson,
  Davis, Laurie, Neumann, Norton, and Roe}{Woodruff et~al\mbox{.}}{2014}]%
        {Woodruff14}
\bibfield{author}{\bibinfo{person}{Jonathan Woodruff},
  \bibinfo{person}{Robert~N.M. Watson}, \bibinfo{person}{David Chisnall},
  \bibinfo{person}{Simon~W. Moore}, \bibinfo{person}{Jonathan Anderson},
  \bibinfo{person}{Brooks Davis}, \bibinfo{person}{Ben Laurie},
  \bibinfo{person}{Peter~G. Neumann}, \bibinfo{person}{Robert Norton}, {and}
  \bibinfo{person}{Michael Roe}.} \bibinfo{year}{2014}\natexlab{}.
\newblock \showarticletitle{The CHERI Capability Model: Revisiting RISC in an
  Age of Risk}. In \bibinfo{booktitle}{{\em Proceeding of the 41st Annual
  International Symposium on Computer Architecuture}} {\em
  (\bibinfo{series}{ISCA '14})}. \bibinfo{publisher}{IEEE Press},
  \bibinfo{address}{Piscataway, NJ, USA}, \bibinfo{pages}{457--468}.
\newblock
\showISBNx{978-1-4799-4394-4}
\showURL{%
\url{http://dl.acm.org/citation.cfm?id=2665671.2665740}}


\end{thebibliography}

\appendix
\newpage

\lstset{style=customc}

\begin{figure*}[h]
\myappxsection{EXCERPT FROM SREPLACE}
\label{appx:sreplace}

\raggedright Listing~\ref{lst:sreplace} shows an excerpt from the vulnerable \texttt{sreplace()} function in ProFTPD.
\begin{minipage}[t]{\textwidth}
\begin{lstlisting}[
  texcl,
  lineskip=-0.7ex,
  label=lst:sreplace,
  caption={\ifabridged\footnotesize\fi Excerpt from \texttt{sreplace()} function in \texttt{proftpd/src/support.c}~\cite{Hu16}. The \texttt{src} pointer is set to the next character of the input string containing replacement patterns (\dOne). When input does not match a replacement pattern, the character is copied verbatim to the output buffer (\dTwo). The preceding bounds check is off-by-one (\dThree), allowing the null-terminator to be overwritten. During the immediately following iteration of the \texttt{while} loop, \texttt{strlen(bpuf)} will exceed the size of \texttt{blen}, resulting in an integer underflow of the \texttt{n} parameter to \texttt{sstrncpy()} (\dFour), allowing the attacker to overwrite the local variables on the stack and gain control of \texttt{sreplace()}}.
]
char *sreplace(pool *p, char *s, ...) {
  char *src = s;   // \dOne\ pointer to next unprocessed character in input buffer parameter
  char *cp;        // pointer to next free entry in buf
  char **mptr;     // pointer to replacement pattern
  char **rptr;     // pointer to replacement string
  char *marr[33];  // replacement pattern string array
  char *rarr[33];  // replacement string array
  *pbuf = NULL;    // pointer to beginning of buf
  
  // destination buffer
  char buf[PR_TUNABLE_PATH_MAX] = {'\0'};

  // unsigned sizes of pattern, string and buf
  size_t mlen = 0, rlen = 0, blen;

  cp = pbuf = buf;
  blen = sizeof(buf);

  while (*src) {   // until input exhausted...
    // ...match substring against all patterns
    for (mptr = marr, rptr = rarr;
         *mptr;    // null indicates end of array
         mptr++, rptr++) {
      mlen = strlen(*mptr);    // length of pattern
      rlen = strlen(*rptr);    // length of string
      if (strncmp(src, *mptr, mlen) == 0) {
        sstrncpy(              // copy replacement to buffer
          cp, *rptr,           // dest, src
          // \dFour\ integer underflow when blen less than length of pbuf string
          blen - strlen(pbuf)  // n bytes to copy
        );
      }
    }

    if (!*mptr) {                    // no pattern matched
      if ((cp - pbuf + 1) > blen) {  // \dThree\ off-by-one bounds-check
        cp = pbuf + blen - 1;
      }
      *cp++ = *src++;                // \dTwo\ copy character verbatim
    }
  }
}
\end{lstlisting}
\end{minipage}

\myappxsection{EXCERPT FROM PR\_STRTIME}
\label{appx:pr_strtime}

Listing~\ref{lst:pr_strtime} shows an excerpt from the \texttt{pr\_strtime()} function in ProFTPD~\cite{Hu16}.

\lstset{basicstyle=\tiny,escapechar=@,style=customc}
\begin{minipage}{\textwidth}
\begin{lstlisting}[
  texcl,
  label=lst:pr_strtime,
  caption={\ifabridged\footnotesize\fi Excerpt of ProFTPd's \texttt{pr\_strtime()} function with \texttt{mons} array. \texttt{pr\_strtime()} indexes \texttt{mons} by month number, copies the corresponding string literal to the static output buffer, and returns a pointer to the output buffer.}]
const char *pr_strtime(time_t t) {
  static char buf[30];                  // output buffer
  static char *mons[] = { "Jan", "Feb", // $\ldots$ }
\end{lstlisting}
\end{minipage}
\end{figure*}

\clearpage

\myappxsection{OTHER KNOWN DOP ATTACKS}
\label{appx:otherdop}

\subsection{DOP attacks against ProFTPD}
\label{appx:proftpd-dop}

In addition to the DOP program that bypasses randomization defenses discussed in Section~\ref{subsec:security_considerations}, Hu \emph{et al.}~\cite{Hu16} present two additional end-to-end attacks against ProFTPD. Both attacks build on the idea of leveraging DOP to corrupt relocation information maintained by the Linux runtime linker. The runtime linker maintains a \texttt{link\_map} structure for each loaded ELF object. The \texttt{link\_map} structure contains the name of the corresponding ELF object, the base address at which the object is loaded, and the virtual address of all the ELF object's dynamic metadata tables. 

In the DOP attack, the memory corruption vulnerability in \texttt{sreplace()} is used to inject specially crafted symbol and relocation metadata into the program's data section. The linked list of \texttt{link\_map} structures is then corrupted to include the malicious relocation metadata. The malicious relocation metadata is consumed by the POSIX \texttt{dlopen()} function. \texttt{dlopen()} is responsible for patching the relocated addresses before execution of the loaded module. In order to do so, \texttt{dlopen()} has the ability to modify arbitrary memory locations, even code pages or read-only data sections. ProFTPD invokes \texttt{dlopen()} in its PAM module to dynamically load libraries. By corrupting a 
field in a global static data structure used by PAM subsystem to keep track of loaded modules, the attacker can trigger the invocation of \texttt{dlopen()}. As \texttt{dlopen()} is invoked, it will process the malicious relocation metadata and trigger an arbitrary memory corruption. This gives the attacker a powerful DOP gadget capable of bypassing defenses based on non-writable code or read-only data.

\ABBRNAME prevents the attacker from escalating the \texttt{sreplace()} assignment gadget into a powerful \texttt{dlopen()}-based assignment in two ways. First, \ABBRNAME can prevent \texttt{sreplace()} from tampering with the \texttt{link\_map} structure. Second, it can prevent \texttt{sreplace()} from modifying the data structures belonging to the PAM subsystem.

\subsection{DOP attacks against Wireshark}
\label{appx:wireshark-dop}

\begin{raggedright}

Hu \emph{et al.}~\cite{Hu16} also describe a DOP attack against the Wireshark~\footnote{https://www.wireshark.org/} network protocol analyzer. The attack targets a stack buffer overflow vulnerability CVE-2014-2299 ~\footnote{https://cve.mitre.org/cgi-bin/cvename.cgi?name=CVE-2014-2299} in the \texttt{packet\_list\_dissect\_and\_cache\_record()}, as shown in Listing~\ref{lst:wireshark}. A pointer to the local stack buffer \texttt{pd} in \texttt{packet\_list\_dissect\_and\_cache\_record()} is passed to \texttt{cf\_read\_frame\_r()} which overflows it (\dOne). This allows the attacker to control the contents of stack variables \texttt{col} and \texttt{*cinfo} (\dTwo). In order to chain gadgets found in \texttt{packet\_list\_change\_record()} (\dThree, \dFour) together, the attacker must corrupt the loop condition \texttt{cell\_list} (\dFive) in the parent function \texttt{gtk\_tree\_view\_column\_cell\_set\_cell\_data()} to ensure that the loop executes infinitely. To achieve this, the \texttt{cell\_list} is directed to a malicious linked-list in the malicious payload.

\ABBRNAME prevents the initial buffer overflow by ensuring that the  \texttt{packet\_list\_dissect\_and\_cache\_record()} function is only delegates access the \texttt{pd} array. Thus \texttt{cf\_read\_frame\_r()} is prevented from accessing variables in its caller's stack frame. Similarly, ensuring the availability of the gadget dispatcher requires an out-of-scope access from 
\texttt{packet\_list\_change\_record()} to the \texttt{cell\_list} variable local to \texttt{gtk\_tree\_view\_column\_cell\_set\_cell\_data()}. \ABBRNAME can thus prevent the DOP program establishing a reliable gadget dispatcher for the attack.  

\end{raggedright}

\begin{minipage}{\mylistingwidth}
\begin{lstlisting}[ 
  texcl,
  label=lst:wireshark,
  caption={\ifabridged\footnotesize\fi Excerpt from \texttt{wireshark/ui/gtk/packet\_list.c}~\cite{Hu16}}.]

void packet_list_dissect_and_cache_record(
  PacketList *packet_list, /* ... */) {
  gint col;
  column_info *cinfo;     // \dTwo
  guint8 pd[WTAP_MAX_PACKET_SIZE];
  
  // \dOne\ initial buffer overflow 
  cf_read_frame_r(/* ... */, fdata,
                  /* ... */, pd);
  
  packet_list_change_record(packet_list,
                            /* ... */,
                            col, cinfo);
}

void packet_list_change_record(
    PacketList * packet_list,
    PacketListRecord *record,
    gint col,
    column_info *cinfo) {
  record = packet_list->physical_rows[row];
  
  // \dThree\ assignment/load/store gadget
  record->col_text[col] =
  (gchar *) cinfo->col_data[col];
  
  // \dFour\ conditional assignment gadget
  if (!record->col_text_len[col])
    ++packet_list->const_strings;
}

void gtk_tree_view_column_cell_set_cell_data(
    GtkTreeViewColumn *self,
    GtkTreeModel* tree_model,
    GtkTreeIter* iter,
    gboolean is_expander,
    gboolean is_expanded);

  /* ... */

  // \dFive\ loop condition corrupted to enable gadget dispatch
  for (cell_list = tree_column->cell_list;
       cell_list;
       cell_list = cell_list->next) {
    /* ... */
    
    // eventually calls vulnerable function
    show_cell_data_func();
    
    /* ... */
}
\end{lstlisting}
\end{minipage}

\newpage

\subsection{DOP attacks against GStreamer}
\label{appx:gstreamer-dop}

Evans~\cite{Evans16} demonstrates an attack against the GStreamer FLIC decoder\footnote{https://gstreamer.freedesktop.org} that exploits a combination of control-flow hijacking and DOP techniques. The attack exploits a decode loop that lacks bounds checks against the output \texttt{frame\_data} buffer, as shown in Listing~\ref{lst:decodeloop}. The exploit can be triggered via a specially crafted media file that causes the decode loop to write past the bounds of the heap-resident buffer. The goal of the attacker is to escalate the buffer overflow into an arbitrary code execution, but they cannot perform a traditional control-flow attack due to the presence of ASLR. The initial memory corruption vulnerability is non-linear, allowing the attacker to skip over heap memory before the write, but only allows the attacker to tamper with memory below the overflowing \texttt{frame\_data} buffer. Additionally the attacker targets program logic that extracts metadata from a media file which only runs the decode loop for two media frames, considerably limiting the initial number of write gadgets possible to chain together.

GStreamer decoders are typically run in their own dedicated threads with their own thread heap. When decoding begins in a newly created decoder thread, the predictability of the initial heap layout allows the attacker to corrupt the \texttt{flxdec} metadata object (Listing~\ref{lst:gstflxdec}) typically allocated in the heap at a predictable offset below the \texttt{frame\_data} buffer used by the decoder. By massaging the \texttt{flxdec} object into a state that keeps the decode loop running beyond the initial two frame window, the attacker obtains a DOP dadget dispatcher used to execute the subsequent DOP payload. The DOP program leverages data-oriented gadgets found in \texttt{flx\_decode\_delta\_fli()} (Listing~\ref{lst:decodeloop}) and \texttt{gst\_flxdec\_chain()} (Listing~\ref{lst:gstchain}) The assignment gadget leverages the \texttt{memcpy()} from \texttt{flxdec->delta\_data} to \texttt{flxdec->frame\_data} (\dOne). By corrupting these pointers in the decode loop, the gadget can be used for arbitrary memory loads and stores. Combining this with the assignment gadget \texttt{flxdec->frame\_data} to \texttt{flxdec->delta\_data} in \texttt{gst\_flxdex\_chain()} (\dTwo) provides a dereference gadget. Together with the assignment gadget (\dThree), these gadgets allow the DOP program to perform a load / add / store during frame processing. With this computational capability the attacker can obtain a code pointer to a known function within the program's code section, add an offset to it to obtain a pointer into the program's global offset table (GOT) which contains relocated pointers to shared library functions. Reading a GOT entry for a shared library function reveals the randomized code pointer. This in turn enables the attacker to mount a traditional control-flow attack on the derandomized code.

GStreamer attack operates by corrupting a heap object which is legitimately within the scope of the decode loop with the initial memory corruption vulnerability. Thus, preventing the GStreamer attack is more challenging compared to the ProFTPD attacks described in Section~\ref{subsec:dop_mitigation} and Appendix~\ref{appx:proftpd-dop}. However, \ABBRNAME can be utilized by the caller to selectively delegate individual fields of the dynamically allocated \texttt{flxdec} in \texttt{flx\_decode\_delta\_fli()} \texttt{gst\_flxdex\_chain}. Furthermore, by applying \ABBRNAME at a block-granularity, the vulnerable decode loop can be constrained to stay within the bounds of the  \texttt{flxdec->delta\_data} and \texttt{flxdec->frame\_data} buffers.

\begin{minipage}{\mylistingwidth}
\begin{lstlisting}[
  texcl,
  label=lst:decodeloop,
caption={\ifabridged\footnotesize\fi Excerpt from vulnerable decode loop in \texttt{gst-plugins-good/gst/flx/gstflxdec.c}~\cite{Evans16}. The attacker controls the contents of the \texttt{data} buffer and can selectively write from \texttt{data} to memory beyond the bounds of \texttt{flxdec->frame\_data} by controlling the \texttt{start\_line}, \texttt{lines}, and \texttt{packets} variables. The data-oriented assignment gadget \dOne\ is reachable by corrupting the \texttt{flxdec} heap object.}
]
// called with dest argument flxdec->frame\_data
flx_decode_delta_fli(GstFlxDec * flxdec,
                     guchar * data,
                     guchar * dest) {
  // \dOne\ data-oriented assignment gadget
  memcpy(dest, flxdec->delta_data, flxdec->size);

  start_line = (data[0] + (data[1] << 8));
  lines = (data[2] + (data[3] << 8));
  data += 4;

  /* start position of delta */
  dest += (flxdec->hdr.width * start_line);
  start_p = dest;

  while (lines--) {
    /* packet count */
    packets = *data++;

    while (packets--) {
      /* skip count */
      dest += *data++;

      /* RLE count */
      count = *data++;

      if (count > 0x7f) {
        /* ... */
      } else {
        /* replicate run */
        while (count--)
        *dest++ = *data++;
  /* ... */

\end{lstlisting}
\end{minipage}

\begin{minipage}{.95\columnwidth}
\begin{lstlisting}[
  texcl,
  label=lst:gstflxdec,
  caption={\ifabridged\footnotesize\fi Excerpt of \texttt{flxdec} struct from \texttt{gst-plugins-good/gst/flx/gstflxdec.h}~\cite{Evans16}.}
]
struct _GstFlxDec {
  GstElement element;
  GstPad *sinkpad;
  GstPad *srcpad;  // pointer to
                   // gst\_flxdec\_chain()
  gboolean active, new_meta;
  guint8 *delta_data, *frame_data;
  GstAdapter *adapter;
  gulong size;
  GstFlxDecState state;
  gint64 frame_time;
  gint64 next_time;
  gint64 duration;
  FlxColorSpaceConverter *converter;
  FlxHeader hdr;
};
\end{lstlisting}
\end{minipage}

\begin{minipage}{\mylistingwidth}
\begin{lstlisting}[
  texcl,
  label=lst:gstchain,
caption={\ifabridged\footnotesize\fi Excerpt from \texttt{gst\_flxdec\_chain()} in \texttt{gst-plugins-good/gst/flx/gstflxdec.c}~\cite{Evans16}. The data-oriented gadgets \dTwo\ and \dThree\ are reachable by corruping the \texttt{flxdec} heap object.}
]
gst_flxdec_chain(/* ... */) {
  
  flx_decode_chunks(
      flxdec,
      ((FlxFrameType *) chunk)->chunks,
      chunk + FlxFrameTypeSize,
      flxdec->frame_data);

  // \dTwo\ data oriented assignment gadget
  memcpy(flxdec->delta_data,
         flxdec->frame_data,
         flxdec->size);

  gst_buffer_map(out, &map, GST_MAP_WRITE);
  /* convert current frame. */
  flx_colorspace_convert(
      flxdec->converter,
      flxdec->frame_data,
      map.data);

  gst_buffer_unmap (out, &map);

  GST_BUFFER_TIMESTAMP (out) = flxdec->next_time;
  // \dThree\ data oriented addition gadget
  flxdec->next_time += flxdec->frame_time;

\end{lstlisting}
\end{minipage}

\myappxsection{STRICT DELEGATION}
\label{appx:strict_delegation}
The \emph{lax delegation} semantics for \texttt{srsub} and \texttt{srdlg}(\texttt{m}) (Section~\ref{sec:instructions}) are not ideal in cases where a memory error in function allows an attacker to corrupt a pointer subsequently delegated to a callee function that operates with higher RSE privileges than the caller (i.e., situations where the caller itself created the pointer, but does not dereference it on its own). With lax delegation, the callee may become the target of a confused deputy attack unless it validates the pointer delegated to it prior to dereferencing. 
Addressing the lack of such input validation is out-of-scope for \SHORTNAME, as is protecting data which is legitimately in the scope of a vulnerable function. 
However a small modification to delegation semantics can prevent these confused deputy situations for a specific class of programs in which the caller's privileges are always a superset of the callee's privileges with respect to a delegated pointer.
This is achieved by implementing \emph{strict delegation}, whereby the behaviour of \texttt{srsub} and \texttt{srdlg}(\texttt{m}) is modified to generate a hardware fault if a matching SRS entry for the specified sub-entry is not found. This ensures that delegated pointers always target a subset of the the caller's RSE priviledges. In addition, \texttt{srsub} and \texttt{srdlg}(\texttt{m}) must be modified to allow (ignore) the special case of delegating a null-pointer. However, determining whether a program is suitable for strict delegation may require dynamic data-flow program analysis.

\myappxsection{RETURN STATE PROTECTION}
\label{appx:ra-protection}

For ease of understanding, our examples so far have described a slightly simplified version of \ABBRNAME instrumentation in which a single SRS entry is created for a function's entire stack frame. In this scenario, memory errors in the main body of the function could allow an attacker to modify the function's return address and other return state information stored in the beginning of the stack frame. However, under normal circumstances, this return state information should only be accessed by the function prologue and epilogue, not by code in the main body of the function. In our actual \ABBRNAME design, we enforce this separation by placing the function's prologue and epilogue into a different execution context from the function's main body. The return state information is thus covered by a different SRS entry from the local stack variables.

Listing~\ref{lst:ra-protection} demonstrates how our full \ABBRNAME instrumentation protects the return address and saved frame pointer against modification by potentially vulnerable code within the function body. Since the prologue and epilogue surround the function's main body, we instrument these to execute in \emph{two distinct execution contexts}. The prologue (\dOne) and epilogue (\dFive) \emph{share their execution context} and SRS frame, and can thus access the same area of the stack. They can also access argument objects passed by reference and delegated by the caller, and delegate these forward to the function body (\dTwo). Once the execution context of the function body has been entered (line 10) an additional SRS entry is created for local stack variables (\dThree), before the execution of the main function body begins (\dFour). Exiting the function body's execution context returns to the execution context set up by the prologue, whilst control flow proceeds to the epilogue (\dFive).
This protects the return state information during the execution of the function body, and thus mitigates attacks that require modification of this control-flow information (e.g., ROP).

\begin{lstlisting}[
  float,
  texcl,
  style=customasm,
  label=lst:ra-protection,
  caption={\ifabridged\footnotesize\fi Instrumented function with return state protection. The function prologue and epilogue share the same execution context.}
]
; \dOne\ function prologue
srdda   -8(sp),sp  ; create SRS entry for return state 
addi    sp,sp,-32  ; decrement stack pointer
sw      ra,24(sp)  ; store return address at sp+24
sw      s0,20(sp)  ; store frame pointer at sp+20
addi    s0,sp,32   ; update frame pointer
; \dTwo\ argument delegation from prologue to body
srdlg   a0         ; 1st argument pointer
srdlg   a1         ; 2nd argument pointer
sbent              ; enter body context
; \dThree\ create SRS entry for local variables
sradd   sp,23(sp)
; \dFour\ function body
; \ldots
srdlg   a0         ; delegate returned object
sbxit              ; exit into epilogue context
; \dFive\ function epilogue
lw      ra,24(sp)  ; load return addr. from stack
lw      s0,20(sp)  ; load frame pointer from stack
addi    sp,sp,32   ; increment stack pointer
srdlg   a0         ; delegate returned object
sbxit              ; exit function context
ret                ; return to caller
\end{lstlisting}

\myappxsection{EXTENDING \SHORTNAME}
\label{appx:extensions}

As stated in section~\ref{sec:assumptions}, in this work we restrict our attention to single-threaded C programs. In this appendix, we outline the challenges in extending \SHORTNAME to support \emph{multi-threaded programs}, \emph{Symmetric Multi-Processing} (SMP), and C++, and we suggest possible solutions to these challenges. The discussion here is not exhaustive, and the implementation and evaluation of these extensions is beyond the scope of this work. 

\myparagraph{Multi-threading} The current implementation of \SHORTNAME maintains only a single SRS, and hence does support multiple concurrent execution contexts. However, when multiple processes or threads are executed concurrently (e.g., interleaved by the CPU), each must be associated with a distinct \SHORTNAME SRS. To facilitate this, \SHORTNAME must provide the means to store the full SRS state of a thread when the thread is pre-empted, and restore the thread's SRS state when it is scheduled. The system must allocate separate storage for each thread's SRS and the \SHORTNAME hardware must maintain a pointer to the current SRS e.g., in a dedicated register. The system scheduler must update this SRS pointer register when switching threads. Access controls should be put in place to prevent unprivileged threads from tampering with any of the stored SRS states, including their own, or with the SRS pointer register. Privileged software must also have the ability to flush all records in the active and spare banks into the SRS in memory during thread scheduling, to ensure the coherency of the stored SRS. 

Threads that co-operate on a computing task may need to share data amongst each other e.g., by passing pointers to shared memory areas. To facilitate this, threads may need to delegate storage region entries to one another. Currently, \SHORTNAME does not provide means of delegation across multiple SRSs. In order to enable delegated access to shared memory areas, \SHORTNAME must be extended with a \emph{messagebox} facility that allows the delegating thread to identify a recipient thread and mark a storage region entry for delegation to the recipient.

\myparagraph{Symmetric Multi-Processing} In addition to the enhancements to support multi-threading, SMP systems require additional considerations to allow \SHORTNAME to maintain SRSs across multiple processors or cores. Most importantly, the \SHORTNAME active, spare, and cache banks must be duplicated for each distinct core. The SRS should be maintained in memory that is accessible to each core, and the messagebox facility must be extended to allow delegations across cores. 

\myparagraph{C++} The most important factor in terms of extending \SHORTNAME to support the C++ runtime is enabling C++ exception handling. Exceptions provide a way to transfer control from one execution context to another, possible separated by several links in the call chain. The exception handling facility has the means to unwind the call stack to the stack frame of the execution context that handles the exception. In \SHORTNAME enabled programs, the exception handling facility must also unwind the SRS to the corresponding SRS frame belonging to the exception context. In the current implementation of \SHORTNAME, the only way to achieve this is via the execution of \texttt{sbxit}. While it would be possible for the exception handling facility to issue multiple consecutive  \texttt{sbxit} instructions until the correct SRS frame has been restored, this is likely to cause undesirable performance overhead due to the associated management operations on the active, spare, and cache banks. Moreover, many of these operations are unnecessary, as the result of the management operations is immediately discarded on the next \texttt{sbxit}. To improve the efficiency of exception handling, \SHORTNAME could be extended with a facility that would enable fast unwinding of the SRS stack to a previous state. A similar facility would also improve the implementation of a \SHORTNAME aware C \texttt{setjmp} / \texttt{longjmp} API.\footnote{http://man7.org/linux/man-pages/man3/setjmp.3.html} 

\myappxsection{COREMARK CONFIGURATION}
\label{appx:coremark_config}

\begin{table*}[t]
  \caption{\ifabridged\footnotesize\fi Breakdown of \SHORTNAME performance overhead by instruction for one iteration of CoreMark validation and performance runs.}
  \label{tbl:coremark_overhead}
  \begin{center}
  \resizebox{.8\textwidth}{!}{    \begin{tabular}{ l | r r r r | r r r r }
      \multicolumn{1}{c}{} & \multicolumn{4}{c}{CoreMark} & \multicolumn{4}{c}{CoreMark} \\
      \multicolumn{1}{c}{} & \multicolumn{4}{c}{(validation run)} & \multicolumn{4}{c}{(performance run)} \\
      \hline
      & \textbf{\# instr} & \textbf{\# cycles} & \textbf{\# stalls} & \textbf{overhead}$^{\mathrm{a}}$
      & \textbf{\# instr} & \textbf{\# cycles} & \textbf{\# stalls} & \textbf{overhead}$^{\mathrm{b}}$ \\
      \hline
        \texttt{sbent}  &  7095 & 31541 & 24446 & 2.00\% &  7157 & 32288 & 25132 & 2.04\%\\
        \texttt{sbxit}  &  7095 & 20288 & 13193 & 1.29\% &  7157 & 20635 & 13478 & 1.31\%\\
        \texttt{sradd}  &  1004 &  1004 & -     & 0.01\% &  1024 &  1024 & -     & 0.01\%\\
        \texttt{srdda}  &  7094 &  7094 & -     & 0.45\% &  7156 &  7156 & -     & 0.45\%\\
        \texttt{srdlg}  &  4327 &  8717 &  4390 & 0.55\% &  4323 &  8944 &  4621 & 0.56\%\\
        \texttt{srdsub} &  2214 &  3424 &  1210 & 0.22\% &  2254 &  3484 &  1230 & 0.22\%\\
        Other           &       &  5353 & -     & 0.34\% &       & 5192 & -      & 0.32\%\\
        \hline
        \textbf{\SHORTNAME} & \multirow{2}{*}{\textbf{28829}}                                & \multirow{2}{*}{\textbf{77421}}                                & \multirow{2}{*}{\textbf{43239}}                                & \multirow{2}{*}{\textbf{4.9\%}}                                & \multirow{2}{*}{\textbf{29071}}                                & \multirow{2}{*}{\textbf{78723}}                                & \multirow{2}{*}{\textbf{44461}}                                & \multirow{2}{*}{\textbf{4.9\%}} \\         \textbf{total} & & & & & & \\
      \hline
      \multicolumn{9}{l}{\footnotesize\raggedright$^{\mathrm{a}}$Calculations based on uninstrumented CoreMark validation run (1578757~cycles).} \\
      \multicolumn{9}{l}{\footnotesize\raggedright$^{\mathrm{b}}$Calculations based on uninstrumented CoreMark performance run (1581062~cycles).} \\
    \end{tabular}}
  \end{center}
\end{table*}

\begin{table}[t]
  \caption{\ifabridged\footnotesize\fi \SHORTNAME performance overhead for different numbers of CoreMark iterations measured on Xilinx Zynq-7020 ZedBoard. \emph{Ticks} indicate the number of timer ticks measured.}
  \label{tbl:coremark_iterations_on_fpga}
  \begin{center}
  \resizebox{\columnwidth}{!}{    \begin{tabular}{ l | r | r r }
      & CoreMark & \multicolumn{2}{c}{CoreMark} \\
      & (uninstrumented) & \multicolumn{2}{c}{(instrumented)} \\
      \hline
      \textbf{Iterations} & \textbf{\# ticks}      & \textbf{\# ticks} & \textbf{overhead} \\
      \hline
        1     &     458187 &     472754 & 3.18\%\\
        100   &   45794484 &   47241900 & 3.16\%\\
        200   &   91580195 &   94473345 & 3.16\%\\
        500   &  228948574 &  236179862 & 3.16\%\\
        1000  &  457906316 &  472369922 & 3.16\%\\
        \hline
        \textbf{Average per} & \multirow{2}{*}{\textbf{457905}} & 
                            \multirow{2}{*}{\textbf{472370}} & 
                            \multirow{2}{*}{\textbf{3.15\%}} \\
        \textbf{iteration}  & & & \\
      \hline
    \end{tabular}}
  \end{center}
\end{table}

The CoreMark benchmark program must be passed three seed values used for initialization of data during the benchmark. The seeds must be input from a source that cannot be determined at compile time to ensure that the compiler cannot pre-compute results to completely optimize away the work intended to be performed during a benchmark. While in principle any seed values could be used, three common sets of seeds have been designated by the CoreMark developers to ensure that results from CoreMark runs remain comparable. The benchmark also contains self-test logic that ensures the correctness of the work performed during the benchmark for the common seeds.

As per industry recommendations\footnotemark we use the \emph{profile run} seeds to obtain the results reported in Section~\ref{subsec:performance_evaluation}. Table~\ref{tbl:coremark_iterations_on_fpga} shows the performance overhead for different numbers of CoreMark iterations measured on the FPGA.
For completeness, we also ran the CoreMark benchmark using the \emph{validation run} and \emph{performance run} seeds to ensure that our instrumentation did not affect the correctness of the benchmark. The results of the validation and performance runs are reported in Table~\ref{tbl:coremark_overhead}. For each run, we fixed the number of CoreMark iterations to one. Because CoreMark was run in a cycle-accurate simulated environment, the results of repeated runs yielded the same cycle counts. For both the validation and performance runs, the number of entries per SRS frame varied between 1 and 120, with a maximum of 11 frames required. The increased SRS utilization was attributed to the increased number of allocations performed by CoreMark's linked list benchmark with the validation and performance seed values.

\footnotetext{\url{http://infocenter.arm.com/help/topic/com.arm.doc.dai0350a/DAI0350A_coremark_benchmarking.pdf}}

Due to area limitations of our FPGA testbed, we were unable to synthesize \SHORTNAME with a frame size of 120 entries. Instead, we ran the validation and performance run benchmarks on both the Spike-based simulator implementation and a cycle-accurate simulation of the hardware implementation (Pulpino SoC extended with \SHORTNAME) in Questa Advanced Simulator\footnote{\url{https://www.mentor.com/products/fv/questa/}}. The cycle-accurate overhead measured for the hardware implementation was at 4.9\%  for the validation and performance runs, in contrast to 3.2\% for the profile run determined on the FPGA.
The principal reason for the increased overhead in Table~\ref{tbl:coremark_overhead} results compared to the profile run was the increased number of stalls (2.8\% overhead for both validation and performance runs). The increase in stalls is a direct result from the increased SRS utilization. Consequently, minimizing the number of stalls is an important consideration for maintaining efficient \SHORTNAME performance in more complex programs. We discuss a possible hardware-design optimization that would reduce the number of stall cycles in Appendix~\ref{appx:performance_optimization}. However, the number of stalls can also be reduced by reducing the required SRS entries for each execution context. This can be achieved by applying \SHORTNAME instrumentation at a finer granularity (Section~\ref{subsec:security_considerations}). Instrumentation at a finer granularity increases the number of required \SHORTNAME context switches, but may provide a net benefit if the distribution of context switches is such that the asynchronous SRS cache update cycles are distributed over a larger portion of machine instructions without causing stalls. Such instrumentation optimization also has the benefit of being adjustable to the performance, and memory access profiles of individual programs without requiring changes to the hardware configuration.    

\myappxsection{Performance Optimization}
\label{appx:performance_optimization}
As discussed in Section~\ref{sec:hardware-impl}, \SHORTNAME context switch instructions \texttt{sbent} and \texttt{sbxit} require at most $N$ additional cycles to transfer the topmost frame between stack and cache in both directions. This implies that the processor must stall if another context switching instruction is encountered within these $N$ cycles. One way to improve the performance at no additional area cost, is to overclock the stack at a frequency higher than that of the processor, i.e., if the processor's operation frequency is $x$ then the stack operation frequency can be set at $2^nx$, where $ 1< n \leqslant log_2{N}$. This would effectively reduce the number of stall cycles to $N$/$2^n$. Note that this comes at the cost of increased power consumption and might not be feasible if the processor is already operating at a high clock frequency.

\end{document}